\documentclass[hyper,letterpaper,notoc]{JHEP3}
\pdfoutput=1
\usepackage{amsmath}
\usepackage{graphicx}% Include figure files
\usepackage[config=altsf]{subfig}
\usepackage{feynmf}
\usepackage{dsfont}
\usepackage{array}
\usepackage{slashed}
\usepackage{afterpage}
\usepackage{appendix}
\usepackage{multirow}

\title{Direct Detection Constraints on Dark Matter Event Rates in
Neutrino Telescopes, and Collider Implications}
\author{Prateek Agrawal\\ Department of Physics, University of Maryland,
College Park, MD 20742}
\author{Zackaria Chacko\\Department of Physics, University of Maryland,
College Park, MD 20742}
\author{Can Kilic\\ Department of Physics and Astronomy, Rutgers University,
Piscataway NJ 08854}
\author{Rashmish K. Mishra\\Department of Physics, University of Maryland,
College Park, MD 20742}

\abstract {
Neutrino telescopes are looking to detect neutrinos produced by the 
annihilation of weakly interacting massive particle (WIMP) dark matter in 
the sun. The event rate depends on the dark matter density in the sun, 
which in turn is dictated by the cross section of WIMPs with nucleons. 
This however is bounded by direct detection experiments. We use the 
constraints from these experiments to place model-independent upper bounds 
on the event rates in neutrino telescopes that apply to any elastic dark 
matter model. Since the spin-independent WIMP-nucleon cross section is 
much more tightly constrained than the corresponding spin-dependent cross 
section, the bounds are much stronger in the former case and are
competitive with the current limits from IceCube. If the number of 
events observed in neutrino telescopes exceeds the upper bound 
corresponding to spin-independent interactions, the implication is that 
the cross section of dark matter with nucleons is dominated by 
spin-dependent interactions. In such a scenario the natural dark 
matter candidates are Majorana fermions and real vector bosons, so that
dark matter particles are their own anti-particles.
We show that any such theory that leads to 
observable event rates at current generation neutrino telescopes will in 
general contain new particles charged under the Standard Model gauge 
groups that naturally lie in a mass range that is kinematically accessible 
to the Large Hadron Collider (LHC).
}

\preprint {UMD-PP-10-005 \\ RUNHETC-2010-08}

\begin{document}
\section{Introduction}

It is now well established that dark matter not contained in the
Standard Model (SM) comprises about 80\% of the total matter in the
universe.  However, the masses, spins and quantum numbers of the
particles of which dark matter is composed are not known. One natural
class of dark matter candidates are Weakly Interacting Massive
Particles (WIMPs). These are stable particles with masses of order the
weak scale that have weak scale cross-sections with visible matter.
When the universe cools WIMPs naturally survive as thermal relics with
the right relic abundance to explain observations. Many well-motivated
extensions of the Standard Model contain WIMP dark matter candidates
that have been shown to naturally give rise to the observed amount of
dark matter, for example supersymmetry
\cite{Ellis:1983ew,Griest:1988ma}, extra dimensional theories
\cite{Kolb:1983fm,Servant:2002aq}, little Higgs models
\cite{BirkedalHansen:2003mpa,Hubisz:2004ft, Birkedal:2006fz} and the
left-right twin Higgs model~\cite{Dolle:2007ce}.

How can WIMP dark matter be detected? In general, there are three classes
of experiments which aim to detect WIMPs. These include
 \begin{itemize}
\item 
direct detection experiments, which look for the recoil of nuclei 
after collisions with WIMPs,
\item
indirect detection
experiments, which search for the annihilation products of WIMPs, and
\item
collider experiments, including the Large Hadron Collider (LHC), which 
hope to observe missing energy signatures associated with the direct 
production of WIMPs.
 \end{itemize}

One promising class of indirect detection experiments involves neutrino 
telescopes, such as IceCube, which are looking to detect neutrinos 
produced by the annihilation of dark matter in the sun 
\cite{Silk:1985ax,Srednicki:1986vj}. This idea has been investigated in 
the context of models with supersymmetric dark matter (for example, 
\cite{Bergstrom:1996kp,Bergstrom:1998xh,Barger:2001ur,Barger:2007xf,Bertin:2002ky,Bertin:2002sq,Hooper:2003ka}, 
for a review see \cite{Jungman:1995df}), Kaluza-Klein dark matter 
\cite{Cheng:2002ej,Hooper:2002gs,Dobrescu:2007ec,Blennow:2009ag,Flacke:2009eu}, 
right-handed neutrino dark matter~\cite{Hooper:2005fj}, little Higgs dark 
matter \cite{Perelstein:2006bq}, the inert doublet model 
\cite{Agrawal:2008xz,Andreas:2009hj,Andreas:2009jp} and right-handed 
sneutrino dark matter \cite{Allahverdi:2009se}. The event rate in neutrino 
telescopes depends on the dark matter density in the sun, which in turn 
depends on the cross section for dark matter scattering off of nucleons. 
This however is bounded by recent direct detection experiments. The number 
of events that could be observed in neutrino telescopes is therefore 
highly correlated with the results of direct detection experiments 
\cite{Kamionkowski:1994dp,Halzen:2005ar,Wikstrom:2009kw,Bandyopadhyay:2010zk}.

In this paper we use the constraints from direct detection experiments
to place upper bounds on the possible event rates in neutrino
telescopes that apply to any elastic dark matter model. These bounds
can be parameterized as a function of the dark matter mass and of the
branching ratios for WIMP annihilation into various SM final states.
Since the spin-independent WIMP-nucleon scattering cross-section is
much more tightly constrained than the corresponding spin-dependent
cross-section, the bounds are much stronger in the former case. We
find that the bounds corresponding to spin-independent interactions
currently stand at about a thousand events per year per km$^2$ for
annihilation directly to neutrinos, and at a few hundred events per
year per km$^2$ for annihilation into any other SM final states. For
most final states, this is stronger than the current limits from
IceCube \cite{Abbasi:2009uz,Abbasi:2009vg}. Furthermore, the upper
bound on the event rate will go down to a handful of events a year if
the direct detection limits improve by two orders of magnitude as
expected. This is below the expected background from atmospheric
neutrinos in neutrino telescopes. The bounds on the event rates
arising from direct detection
constraints on spin-dependent interactions, on the other hand,
currently stand at the level of a few hundred thousand events per
year. Therefore even with expected improvements in the near future,
bounds corresponding to spin-dependent interactions will continue to
be much weaker than the current limits from IceCube.

If the number of events observed in neutrino telescopes exceeds the
upper bounds corresponding to spin-independent interactions, the
implication is that the cross-section of dark matter with nucleons is
dominated by spin-dependent interactions. A model-independent
classification of dark matter candidates that have this property,
based on the spin and quantum numbers of both the WIMP and the
intermediate particle whose exchange mediates the WIMP-nucleon
interaction, has been performed~\cite{Agrawal:2010fh}. From this
study, all theories that lead to elastic spin-dependent WIMP-nucleon
scattering at tree-level through renormalizable interactions have been
identified. It has been established that in such a scenario, the
natural dark matter candidates are Majorana fermions and real vector
bosons, so that the dark matter particle is its own anti-particle.
Scalar and complex vector boson dark matter candidates are disfavored.
Dirac fermions are also disfavored, unless the dark matter particle
carries very specific quantum numbers. Furthermore, it has been shown
that all renormalizable theories with primarily spin-dependent
WIMP-nucleon cross sections at tree-level predict either
\begin{itemize}
  \item new particles at the weak scale with Standard Model quantum 
        numbers, or
  \item a $Z'$ gauge boson with a weak scale mass that serves as a 
        mediator.
\end{itemize}
In this paper we show that in the region of parameter space that is 
accessible to IceCube, these particles naturally lie within the mass range 
that is kinematically accessible to the LHC.

From this discussion we see that there is a very close connection between 
direct detection experiments, dark matter searches in neutrino telescopes 
and the LHC. A null result in the current generation of direct detection 
experiments, in association with a positive signal at IceCube, would 
constitute evidence that dark matter is composed of either Majorana 
fermions or real vector bosons, and therefore that dark particles are 
their own anti-particles. Furthermore, this conclusion would then very 
likely be confirmed by discoveries at the LHC.

The layout of the paper is as follows. In section 2 we give a brief 
overview of the relation between bounds from direct detection experiments 
and event rates at neutrino telescopes, for both spin-independent as well 
as spin-dependent WIMP-nucleon interactions. In section 3, we consider the 
models that give rise to primarily spin-dependent WIMP-nucleon 
interactions, and examine the parameter space of these theories that gives 
rise to an observable signal at IceCube. We show that the masses of the 
new particles naturally lie in a region which is within the energy reach 
of the LHC. We conclude in section 4.

\section{Correlating Direct Detection Experiments and Event Rates in
Neutrino Telescopes} 
\subsection{Bounds on WIMPs from Direct Detection Experiments}

Direct detection experiments are designed to observe the recoil of a
heavy nucleus when a WIMP scatters off it. The typical energy transfer
is of order 10 keV, much smaller than the characteristic nuclear
energy scales. Therefore in such an experiment the WIMP interacts with
the entire nucleus as a single unit, with a net mass, charge and spin.

The cross-section for WIMPs to scatter off different nuclei depends 
strongly on the form of the interactions of the dark matter particle. In 
some cases the WIMP-nucleus cross-section is not sensitive to the spin of 
the nucleus, and we refer to such interactions as spin-independent. In 
other cases the WIMP couples dominantly to the spin of the nucleus, and we 
refer to such interactions as spin-dependent. The bounds on WIMP-nucleus 
scattering from direct detection experiments are generally expressed as 
limits on the WIMP-nucleon cross-section, allowing results across 
experiments using different nuclei to be compared.

The bounds from direct detection experiments on spin-dependent 
interactions are relatively weak since the spin of a typical nucleus is 
either zero or order one, and does not scale with $A$, the number of nucleons. 
Depending upon whether the unpaired nucleon in the nucleus is a proton or 
a neutron, a given experiment will in general place a bound only on the 
spin-dependent WIMP cross-section with either the proton or the neutron. 
On the other hand, spin-independent cross-sections are enhanced by a 
factor of $A^2$, and the bounds on such interactions are correspondingly 
stronger by a factor of order $10^5$. In generating the spin-independent 
limits an implicit assumption is made that the WIMP scattering cross 
section off protons and neutrons is approximately the same, and that there 
are no large cancellations arising from interference between the 
amplitudes for scattering off protons and neutrons.

\begin{figure}[htp]
  \begin{center}
    \subfigure[]{
    \label{fig:directa}
    \includegraphics[scale=0.8]{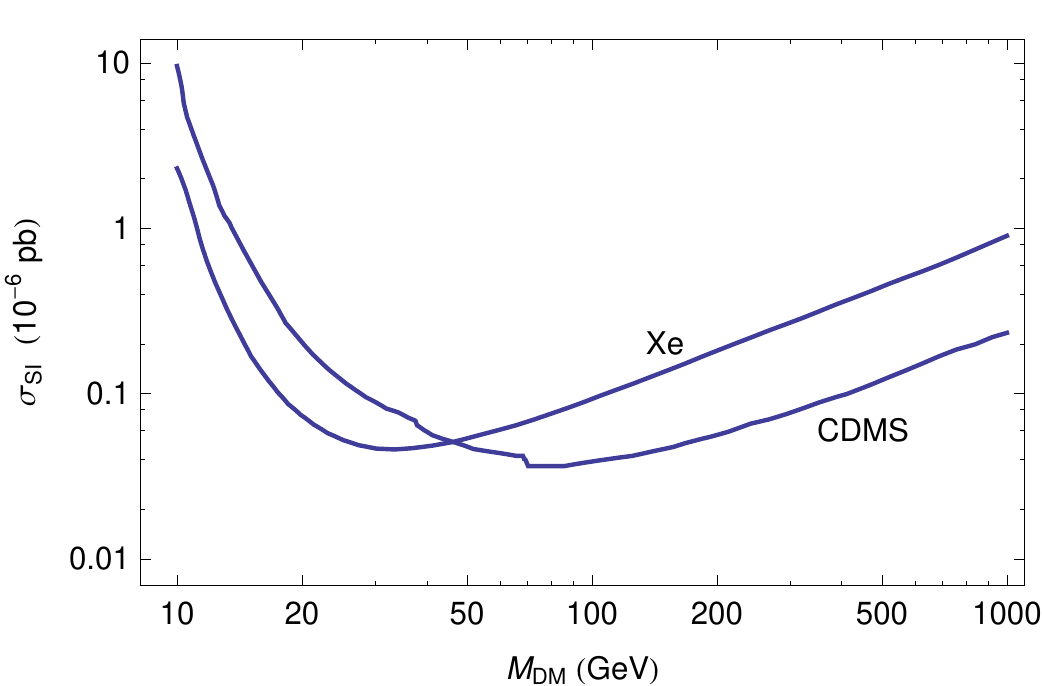}
    }
    \hspace{0.3in}
    \subfloat[]{
    \label{fig:directb}
    \includegraphics[scale=0.8]{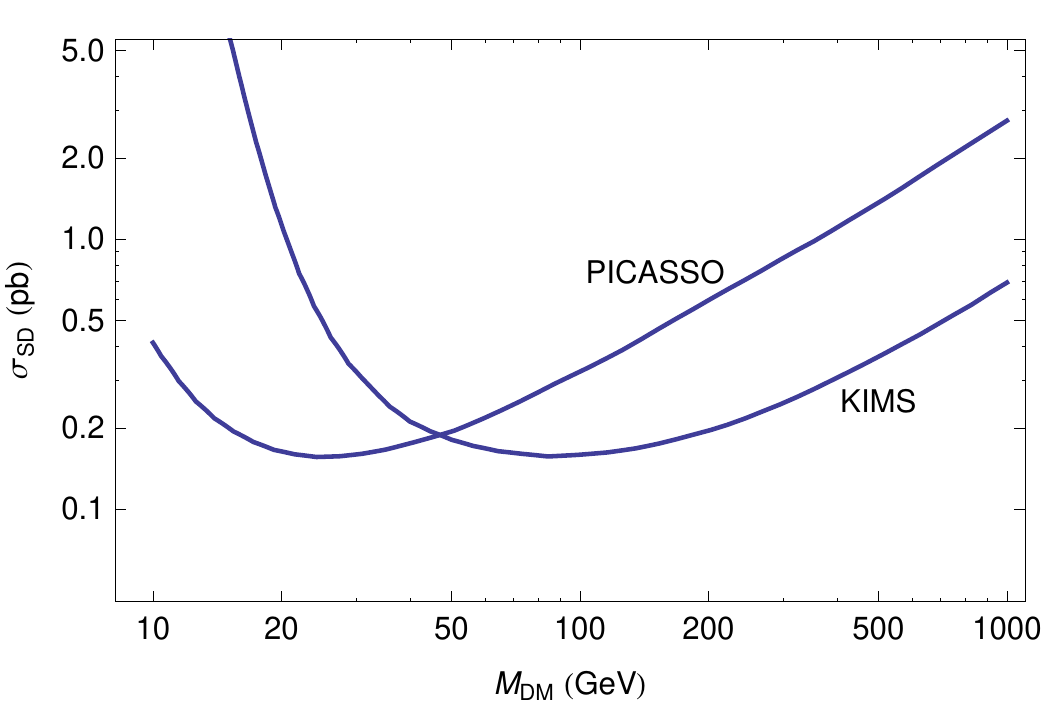}
    }
  \end{center}
  \caption{Current direct detection bounds on the
  a) spin-independent
  dark matter-nucleon cross section and b) the spin-dependent dark
  matter-proton cross section}
  \label{fig:direct}
\end{figure}

In Fig \ref{fig:direct} we show current bounds for the spin-independent 
and spin-dependent WIMP-nucleon scattering cross sections.  As we shall 
explain below, for our purposes only the bound on the spin-dependent 
WIMP-proton cross-section matters, and the bound on spin-dependent 
WIMP-neutron cross-section is irrelevant.  The CDMS \cite{Ahmed:2009zw} 
and Xenon \cite{Angle:2008we} experiments currently put the most stringent 
bounds on spin-independent scattering. For spin-dependent interactions on 
the other hand, the PICASSO~\cite{Archambault:2009sm} and 
KIMS~\cite{Kim:2008zzn} collaborations currently place the tightest bounds 
on the WIMP-proton cross-section. As is clear from the figures, bounds on 
spin-dependent scattering are much weaker, generally by more than five 
orders of magnitude. In the next sub-section we will translate these 
limits on the WIMP-nucleon cross section into an upper bound on WIMP 
capture rates in the sun, and from this into an upper bound on event rates 
in neutrino telescopes.

\subsection{Capture and Annihilation of Dark Matter in the Sun}
\label{sec:nus}

As WIMPs pass through the sun, they are expected to scatter off the
nuclei in the sun and become gravitationally bound. With subsequent
scattering, they eventually accumulate in the center of the sun. These
WIMPs eventually annihilate into various SM final states. The
subsequent decays of these states generally result in high-energy
neutrinos, which can be observed in neutrino telescopes here on earth.

\begin{figure}[htp]
  \begin{center}
    \includegraphics{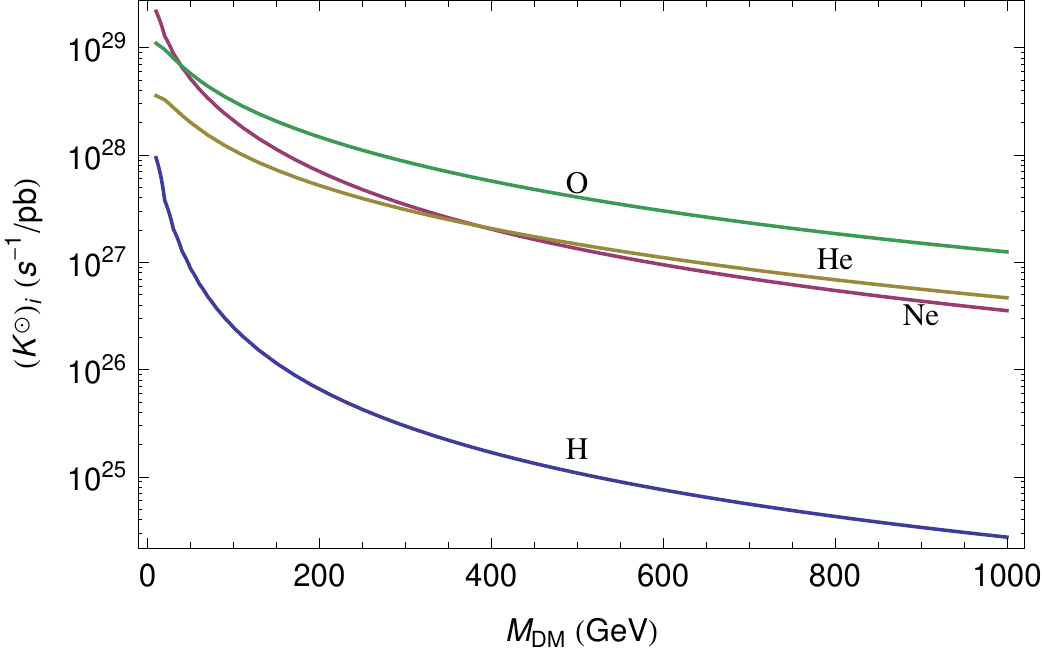}
  \end{center}
  \caption{Contribution of various elements to the spin-independent
  capture rate in the sun}
  \label{fig:captureelem}
\end{figure}

Both spin-independent as well as spin-dependent scattering can lead to the 
capture of dark matter in the sun. Of the nuclei that have a net spin, 
hydrogen is the only one that is present in the sun in significant 
proportions. Other trace elements in the sun generally have no net spin, 
and even when they do, there is no $A^2$ enhancement that can compensate 
for their low density fraction.  Therefore, the only relevant quantity for 
spin-dependent capture is the WIMP-proton cross section. Spin-independent 
capture on the other hand receives contributions from several elements.  
As explained in the previous sub-section, the cross section for 
spin-independent WIMP-nucleus scattering is strongly enhanced by large 
$A$. For WIMP masses significantly larger than the mass of the nucleus, 
the inclusion of kinematic factors causes the WIMP-nucleus cross section 
to scale roughly as $\sim A^4$. Therefore, even though the heavier 
elements in the sun are rarer, this enhancement makes their contribution 
to the capture rate significant.  In particular, as noted in 
\cite{Gould1992}, oxygen plays the most important role in spin-independent 
capture of WIMPs in the sun. We plot the contribution of some 
representative elements to the capture rate in Fig. \ref{fig:captureelem}. 
These numbers were obtained by using the software 
DarkSusy~\cite{Gondolo:2004sc}. We see clearly that heavier elements 
cannot be neglected, and in fact may contribute significantly more to 
capture than either hydrogen or helium.

We present approximate analytical formulae for the capture rate to
understand the dependence on various parameters. The spin-dependent
capture rate in the sun is given by~\cite{Jungman:1995df,Hooper:2005fj},
\begin{align}
  C^\odot_{SD}
  &\simeq 1.3\times 10^{29} \mathrm{s}^{-1}
  \left[ \frac{\rho_{local}}{0.3\ \mathrm{GeV/cm}^3} \right]
  \left[ \frac{270\  \mathrm{km/s}}{\bar{v}_{local}} \right]
  \left[\frac{\sigma_{\mathrm{H,SD}}}
  {1 \mathrm{pb}} \right]
  \left[ \frac{1\text{ GeV}}{m_\mathrm{DM}} \right]
  S(m_{DM}/m_H)
  \\&\equiv K^\odot_{SD} \sigma_{\rm SD} \nonumber
  .
\end{align}
Here, $\sigma_{\text{H,SD}} = \sigma_{\rm SD}$ is the spin-dependent
cross section with hydrogen. The astrophysical parameters
$\rho_{local}$ and $\bar{v}_{local}$ are the local halo density and
root mean square velocity of the dark matter particles, and the
numbers in the formula represent typical values. Finally, since the
capture relies on the final dark matter particle being gravitationally
bound, there is a kinematic suppression $S(m_{DM}/m_H)$ which
depends on the ratio of the WIMP mass to the mass of hydrogen. The
function $S(x)$ approaches 1 for $x\rightarrow 1$, which corresponds
to no suppression.  On the other hand, if the WIMP is much heavier
than hydrogen ($x\rightarrow \infty$), then $S(x) \propto x^{-1}$
while $S(x)\propto x$ for $x\rightarrow 0$.

The spin-independent capture rate, on the other hand, is a sum over
contributions from various elements, and depends on the details of
their relative compositions and distributions in the sun
\begin{align}
  C^\odot_{\rm SI}
  &\simeq 4.8\times 10^{22} \mathrm{ s}^{-1}
  \left[ \frac{\rho_{local}}{0.3\ \mathrm{GeV/cm}^3} \right]
  \left[ \frac{270\  \mathrm{km/s}}{\bar{v}_{local}} \right]
  \left[ \frac{1\text{ GeV}}{m_\mathrm{DM}} \right]
  \nonumber
  \\&\qquad\times
  \sum_i F_i(m_{DM}) f_i \phi_i
  S(m_{DM}/m_{Ni})
  \left[ \frac{1\text{ GeV}}{m_\text{Ni}} \right]
  \left[\frac{\sigma_{\mathrm{N_i, tot}}}
  {10^{-6}\ \mathrm{pb}} \right]
  \\&
  \equiv K^\odot_{SI} \sigma_{\rm SI} \nonumber
  .
\end{align}
Here the approximation that the WIMP-proton and WIMP-neutron cross 
sections are approximately equal has been used to go from the WIMP-nucleus 
cross section ($\sigma_{\mathrm{N_i,tot}}$) to the WIMP-nucleon cross 
section ($\sigma_{\rm SI}$). We now explain the significance of the 
various factors appearing in this equation.  The dimensionless quantity 
$f_i$ corresponds to the relative abundance of the elements in the sun. 
Hydrogen and helium are the most abundant elements in the sun ($f_{H} = 
0.772$ and $f_{He} =0.209$). For all other elements $f_i \sim 10^{-3}$. 
The effect of the different spatial distributions of various elements 
within the sun is accounted for by the dimensionless parameter $\phi_i$. 
It is approximately equal for various elements in the sun $\phi_i\sim 
3.2$, up to corrections of order a few percent. When the dark matter 
particle scatters off heavier nuclei, there is a form factor suppression 
because the WIMP probes only a part of the entire nucleus. The form factor 
$F_i$ starts becoming significant for elements with $A>20$. Finally, since 
the capture relies on the final dark matter particle being gravitationally 
bound, there is again a kinematic suppression $S(m_{DM}/m_N)$ when the 
dark matter mass and the nuclear mass differ. The exact details of this 
calculation may be found in \cite{Jungman:1995df}.

\begin{figure}[h]
  \begin{center}
    \includegraphics{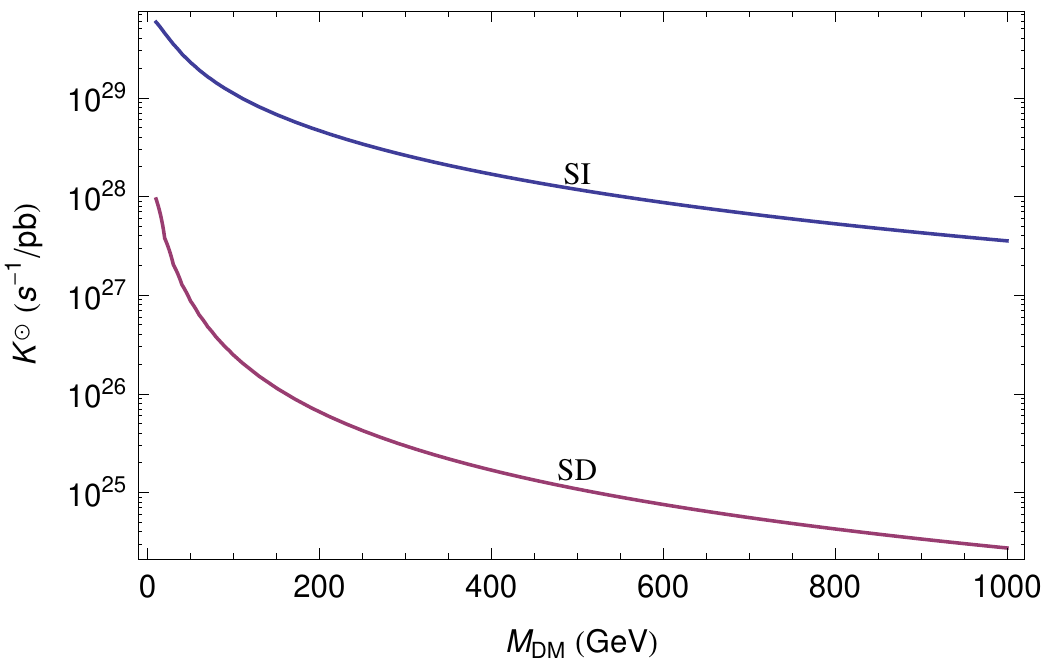}
  \end{center}
  \caption
  {
  Capture ``efficiency'' in the sun for spin-independent and
  spin-dependent scattering
  }
  \label{fig:capture-eff}
\end{figure}

The factors $K^\odot$ are plotted as a function of the dark matter mass in Fig 
\ref{fig:capture-eff}, for both the spin-independent and spin-dependent cases. 
This calculation was performed using DarkSusy~\cite{Gondolo:2004sc}. We can see 
from this that spin-independent WIMP-nucleon scattering is around two to three 
orders of magnitude more efficient in capturing dark matter than the 
corresponding spin-dependent scattering.

The number of dark matter particles ($N$) in the sun is governed by
the simple equation,
\begin{align}
  \dot{N}
  &= C^\odot - 2\,\Gamma_A \nonumber \\
  &= C^\odot - A^\odot N^2
  .
\end{align}
The quantity $\Gamma_A$ which represents the WIMP annihilation rate
depends on the parameter $A^\odot$ which in turn depends on the
annihilation cross section and the effective volume of the core of the
sun, $V_{\mathrm{eff}}$,
\begin{align}
  A^\odot &=
  \frac{\langle \sigma v\rangle}{V_{\mathrm{eff}}}\\
  V_{\mathrm{eff}} &= 1.8\times10^{26} \text{ cm}^3
  \left(  \frac{1000\text{ GeV}}{m_{DM}}\right)^{3/2}
  .
\end{align}
We can solve this equation to obtain the number density today, at
$t_\odot\simeq4.5$ billion years
\begin{align}
  N &= \sqrt{\frac{C^\odot}{A^\odot}}
  \tanh\left(\sqrt{C^\odot A^\odot}t_\odot \right)
  .
\end{align}
If the argument of the $\tanh$ function is large enough, the implication 
is that the process of capture and annihilation in the sun have come into 
equilibrium. This happens if the capture and annihilation cross sections 
are large enough to satisfy $\sqrt{C^\odot A^\odot} t_\odot >1$. In this 
case the annihilation rate is equal to half the capture rate.

\begin{figure}[htp]
  \begin{center}
    \includegraphics[width=0.6\textwidth]{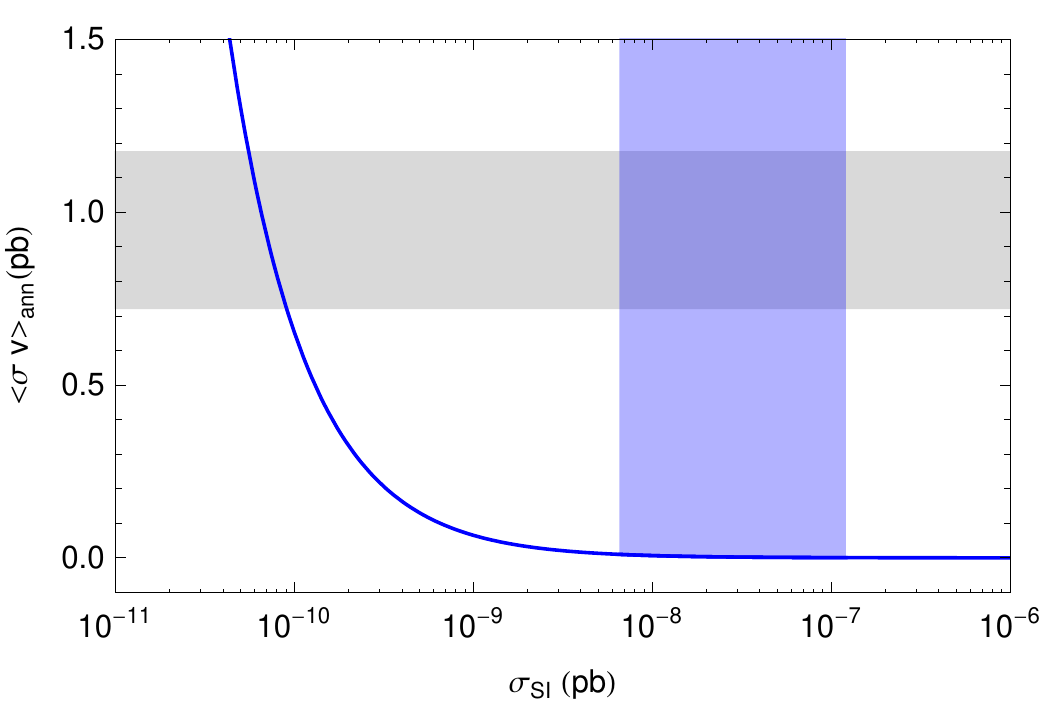}
  \end{center}
  \caption{Equilibrium of dark matter capture/annihilation in the sun
  for mass $m_{DM}=500$ GeV. The region above the curve ensures
  equilibrium. The parameter space to the right of the vertical shaded
  region has already been ruled out by direct detection experiments
  for spin-independent interactions. The region to the left of the
  vertical shaded region will generate fewer than 10 events per year
  per km$^2$ in neutrino telescopes (assuming a $\nu_\tau
  \bar{\nu}_\tau$ final state with $E_{th}=10$ GeV).  The horizontal shaded band
  indicates the annihilation rate required to be consistent with relic
  density observations assuming s-wave annihilation.  }
  \label{fig:con-eq}
\end{figure}
We plot the condition on the scattering and annihilation cross
sections for dark matter to be in equilibrium in Fig \ref{fig:con-eq}. 
It is clear from the figure that for s-wave
annihilation the condition for equilibration is always realized in any
region of parameter space which yields an interesting signal in km$^3$
neutrino telescopes.

For annihilation into any given final state, the total annihilation rate 
$\Gamma$ dictates the total number of high energy neutrinos produced. The 
maximum value of $\Gamma$ depends only on the capture rate $C^\odot$. The 
capture rate in turn is proportional to the WIMP-nucleon scattering cross 
section, which is constrained by direct-detection experiments. Therefore 
the bounds from direct detection lead directly to an upper bound on the 
number of high energy neutrinos produced by WIMP annihilation in the sun 
into any specific SM final state. In the next subsection we will translate 
this into an upper bound on event rates in neutrino telescopes.

\subsection{Upper Bound on Event Rates in Neutrino Telescopes}

Neutrino telescopes are large $\sim {\rm km}^3$ arrays of Cerenkov
detectors placed inside a transparent medium such as ice or water.
Energetic neutrinos from the sun interact with nucleons in the rock
below the detector, and produce muons via charged current
interactions. As these upward going muons propagate through the
detector, they emit Cerenkov radiation, which is detected by the
experiment.

The probability that a muon neutrino of energy $E_\nu$ produces an
observable muon is given by
\begin{align}
  P(E_\nu, E_{th})
  &=
  N_A \; \sigma_{\nu N}(E_\nu) \;
  \langle R(E_\nu;E_{th}) \rangle
  ,
\end{align}
where $\langle R(E_\nu;E_{th}) \rangle$ corresponds to the average
range of the muons produced by neutrinos of energy $E_\nu$ before
their energy drops below $E_{th}$, the energy threshold of the
detector \cite{Dutta:2000hh}. The average is over the fraction of
energy carried by the muon,
\begin{align}
  \langle R(E_\nu; E_{th}) \rangle
  &=
  \frac{1}{\sigma_{\nu N}(E_\nu)}
  \int_0^{1-E_{th}/E_\nu}
  dy \; R(E_\nu(1-y),E_{th})
  \frac{d\sigma_{\nu N}}{dy}(E_\nu, y)
  .
\end{align}
Here $\sigma_{\nu N}(E_{\nu})$ is the neutrino-nucleon charged
current interaction cross section \cite{Gandhi:1995tf} and $(1-y)$ is
the fraction of neutrino energy carried off by the muon. The
definition of the range used here includes the density of standard
rock and is measured in `kilometer water equivalents' ($km w.e.$) or
equivalently, $g/cm^2$.

The flux of muon-events observed is then given
by~\cite{Kopp:2009et,Gandhi:1995tf},
\begin{align}
  \Phi_\mu &=
  \int \frac{d\phi_{\nu}}{dE_{\nu }}
  P(E_\nu, E_{th})
  dE_{\nu}
  \\&=
  \int \int \frac{d\phi_{\nu}}{dE_{\nu}}
  \frac{d\sigma_{\nu N}}{dy}\left( E_{\nu},y \right)
  N_A
  R\left( E_\nu(1-y),E_{th}\right)
  dy \;
  dE_{\nu}
  ,
\end{align}
where $N_A$ is Avogadro's number. Here the differential flux of 
muon neutrinos 
arising from the various final states $F$ is given by
\begin{align}
  \frac{d\phi_{\nu}}{dE_{\nu }}
  &= \sum_F B_F\frac{\Gamma_A}{4\pi d^2}
  \frac{d N^F_{\nu}}{dE_{\nu }}
  ,
\end{align}
where $B_F$ is the branching fraction into the final state $F$,
$\Gamma_A$ is the annihilation rate, $d$ is the distance of the sun
from earth and $d N^F_{\nu}/ d E_{\nu }$ is the energy spectrum of
muon-neutrinos produced in the sun from annihilation into the final
state $F$.

If the threshold energy can be neglected compared to the neutrino
energy, the range of the muon is approximately proportional to the
muon energy. The charged-current cross section turns out to be
proportional to the neutrino energy when the neutrino energy is of
$\mathcal{O}$(TeV). In this case, we can rewrite the above expression,
\begin{align}
  \Phi_\mu
  &=
  \sum_F
  B_F
  \frac{\Gamma_A}{4\pi d^2}
  \int \frac{dN^F_{\nu}}{dE_{\nu}}
  E_\nu^2\;
  dE_{\nu}
  \int_0^1
  N_A
  \frac{1}{E_\nu}
  \frac{d\sigma_{\nu N}}{dy}\left( E_{\nu},y \right)
  \frac{R\left(E_\nu (1-y);0\right)}{ E_\nu}
  dy
  .
\end{align}
With this approximation, the second integral is independent of the
neutrino energy, and can be numerically estimated for a given
detector. Therefore the number of events observed is proportional to the
quantity $\langle Nz^2\rangle$, defined as,
\begin{align}
  \langle Nz^2 \rangle_F (m_{\text{DM}})
  \equiv
  \frac{1}{m_{\text{DM}}^2}
  \int \frac{dN^F_\nu}{dE_{\nu}}
  E_{\nu}^2 \;
  dE_{\nu}
  .
\end{align}
The total rate in the detector is given by the sum of the rate for muons
and the corresponding rate for anti-muons,
\begin{align}
  \Phi_{tot} = \Phi_\mu + \Phi_{\bar{\mu}}
  .
\end{align}
We can then write the total event rate in an empirical form,
\begin{align}
  \Phi_{tot}
  &= (2.54\times 10^{-23} \text{ km}^{-2} \text{ yr}^{-1})
  \left( \frac{\Gamma_A}{s^{-1}} \right)
  \left( \frac{m_{\text{DM}}}{1\text{ GeV}} \right)^2
  \sum_{i=\nu,\bar{\nu}} a_i b_i
  \sum_F B_F \langle N z^2\rangle_F (m_{\text{DM}})
  .
\end{align}
where the $a_i$ are scattering coefficients for neutrinos and
anti-neutrinos, which take values $a_\nu =6.8$ and
$a_{\bar{\nu}}=3.1$, while the $b_i$ are the range coefficients for
muons and anti-muons, with values $b_\nu=0.51$ and
$b_{\bar{\nu}}=0.67$.

\begin{figure}[ht]
  \begin{center}
    \subfloat[]{
    \label{fig:icea}
    \includegraphics[width=0.5\textwidth]{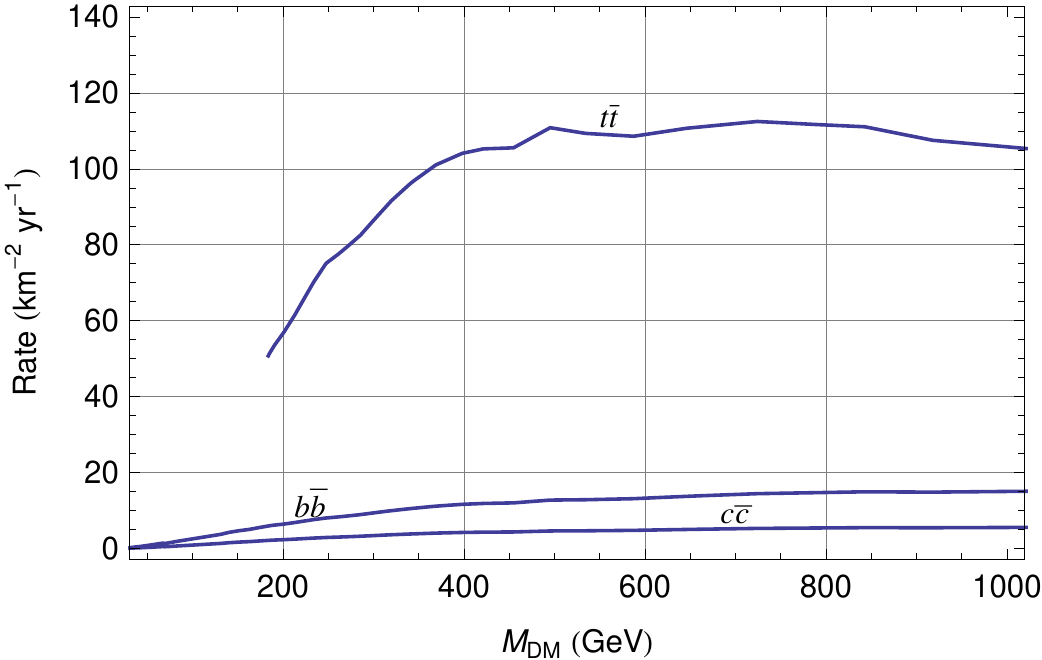}
    }
    \\
    \subfloat[]{
    \label{fig:iceb}
    \includegraphics[width=0.5\textwidth]{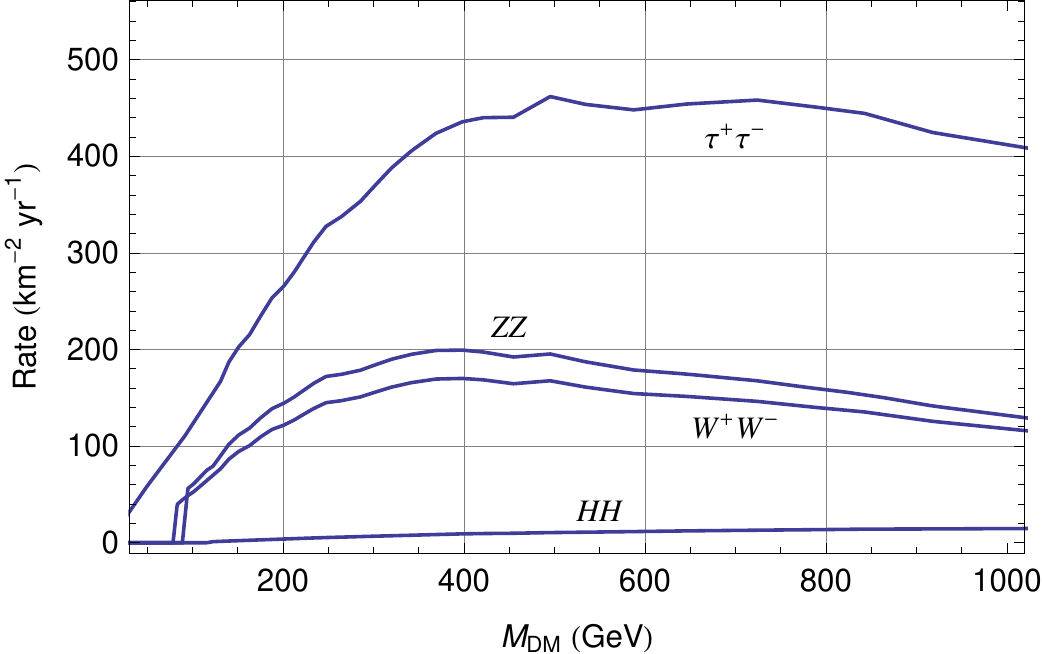}
    }
    \\
    \subfloat[]{
    \label{fig:icec}
    \includegraphics[width=0.5\textwidth]{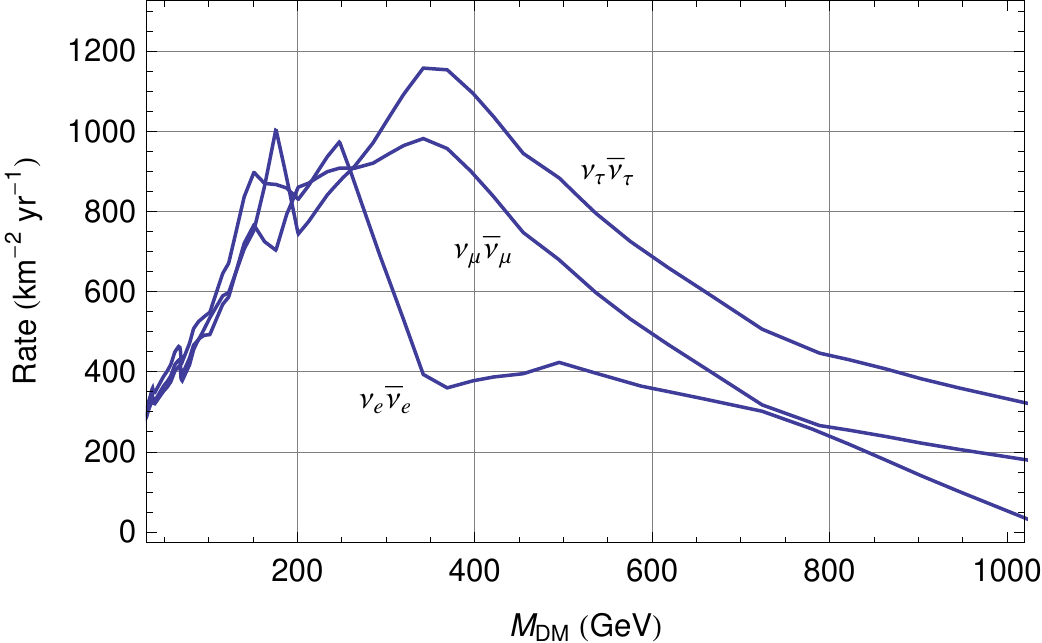}
    }
  \end{center}
  \caption{Maximum event rate in a neutrino detector consistent with
  direct detection bounds from spin-independent
  dark matter for various final states. The threshold energy was
  chosen to be $E_{th}=10$ GeV.}
  \label{fig:bound-si}
\end{figure}

\afterpage{\clearpage}

\begin{figure}[ht]
  \begin{center}
    \subfloat[]{
    \label{fig:icesda}
    \includegraphics[width=0.47\textwidth]{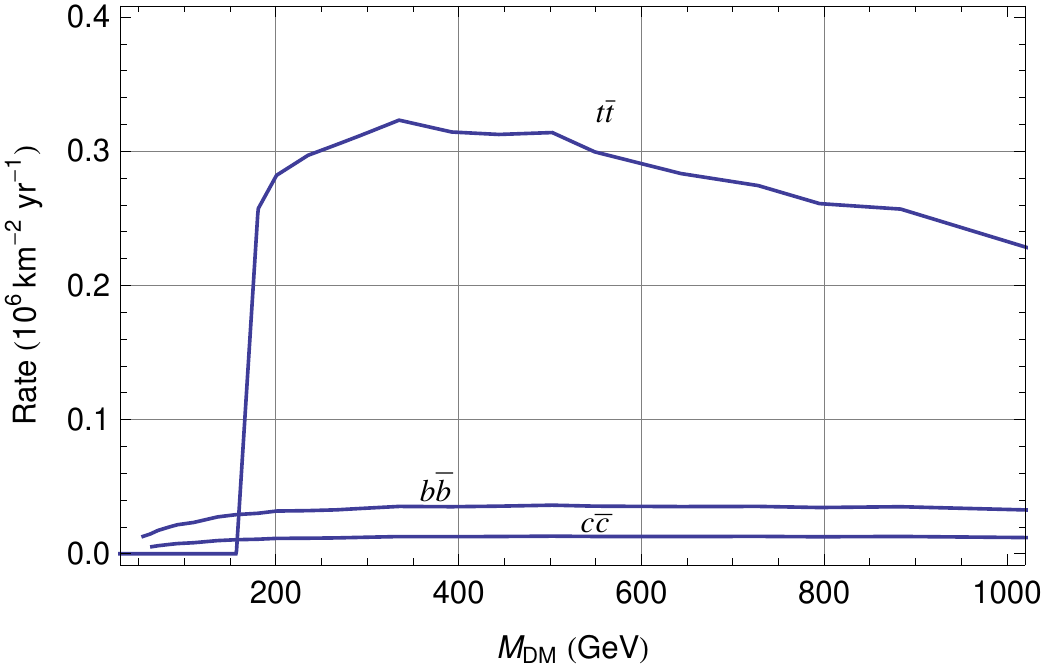}
    }
    \\
    \subfloat[]{
    \label{fig:icesdb}
    \includegraphics[width=0.47\textwidth]{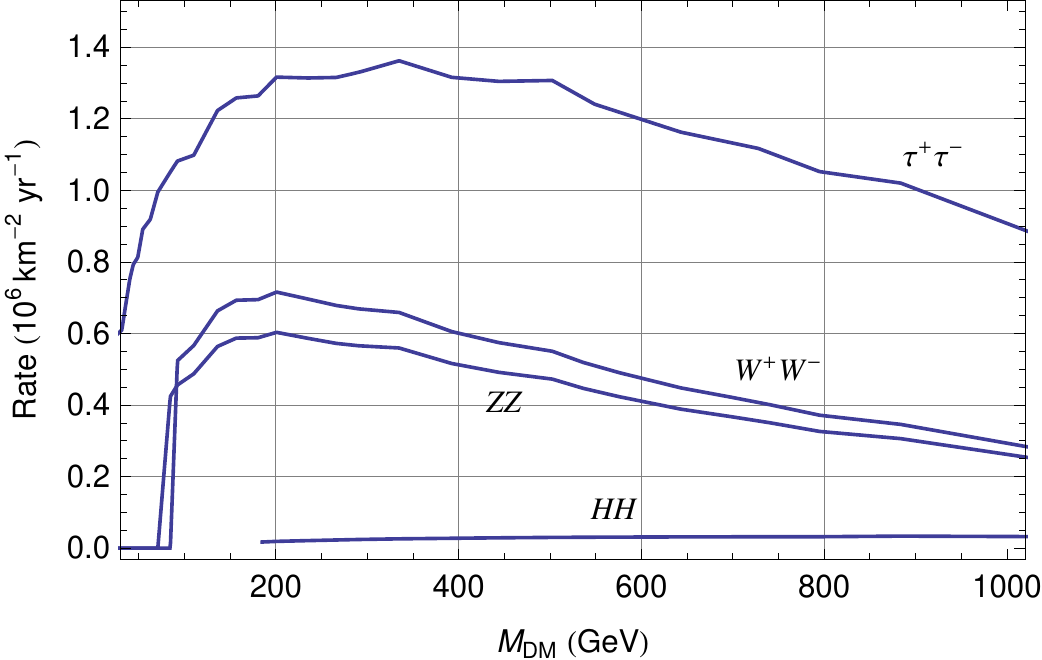}
    }
    \\
    \subfloat[]{
    \label{fig:icesdc}
    \includegraphics[width=0.47\textwidth]{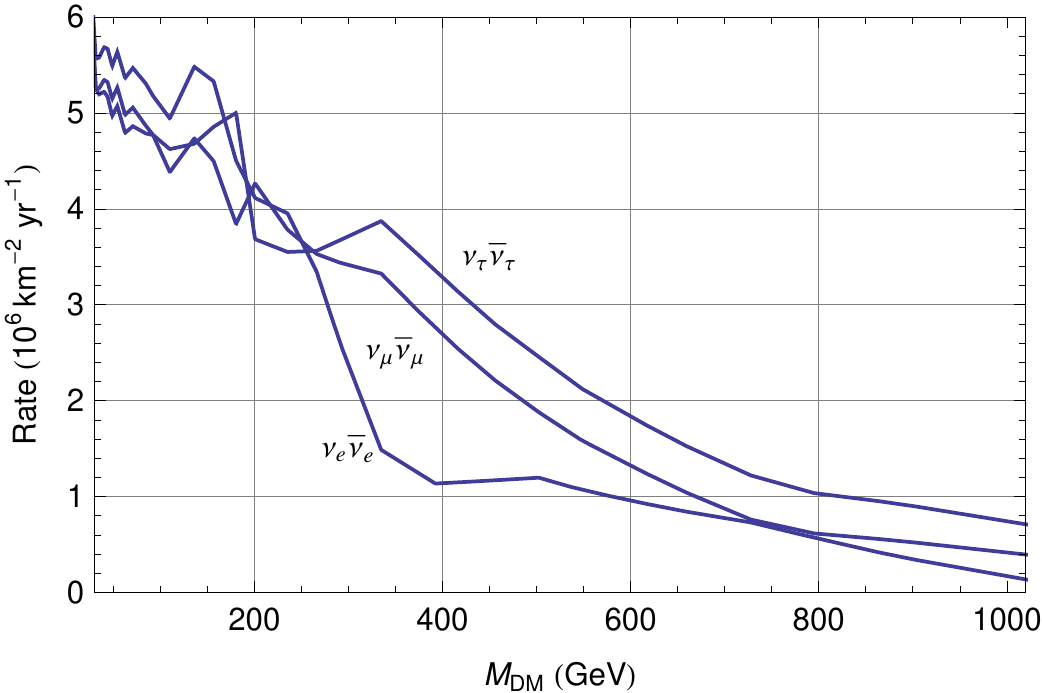}
    }
  \end{center}
  \caption{Maximum event rate in a neutrino detector consistent with
  direct detection bounds from spin-dependent
  dark matter for various final states. The current experimental bounds from
  IceCube and Super-Kamiokande are much stronger. The threshold energy
  was chosen to be $E_{th}=10$ GeV.}
  \label{fig:bound-sd}
\end{figure}

\afterpage{\clearpage} 

In this expression, the effects of neutrino propagation and detector 
thresholds have been neglected. For a relatively heavy WIMP ignoring 
detector threshold effects may be justified because the neutrino (and 
hence the muon) energies are typically higher than the detector threshold 
(for IceCube, $E_{\text{th}} \simeq 10 \text{ GeV})$. Further, since the 
detection rate depends on the second moment of the neutrino spectrum, it 
is dominated by high energy neutrinos. Nevertheless, for a complete 
determination of the signal, the effects of neutrino propagation and the 
detector threshold must be folded in to the above expression. In order to 
estimate their impact on these formulas, we use results from the package 
DarkSusy \cite{Gondolo:2004sc}. We find that these effects are not 
insignificant for certain final states, and can change the answer by as 
much as 50\%. Therefore, the formulae obtained above should only be used 
as a qualitative guide.

We are now in a position to determine the maximum event rate in
neutrino telescopes. We set the WIMP-nucleon scattering cross section
to its upper bound, as determined by the CDMS experiment for
spin-independent scattering and by the KIMS experiment for
spin-dependent scattering. We can
then determine the muon flux at the neutrino telescope and we plot the
result as a function of the WIMP mass for various neutrino-rich two
body SM final states that the WIMPs annihilate into. This gives us the
maximum possible signal rates in neutrino telescopes consistent with
direct detection experiments, assuming that the WIMP has either purely
spin-independent interactions (results plotted in Fig
\ref{fig:bound-si}), or purely spin-dependent interactions (results
plotted in Fig \ref{fig:bound-sd}). In obtaining the results for
annihilation to two Higgs particles, we have assumed a Higgs mass of
120 GeV. As noted above, our formulae do not include (potentially
important) threshold and neutrino propagation effects. Therefore,
we plot the results obtained from DarkSusy.

We see from these figures that models with spin-independent interactions 
do not generate more than about 500 events per year per km$^2$ in neutrino 
detectors, unless annihilation occurs directly to neutrinos, in which case 
the rate can be as large as 1200 events per year per km$^2$. Note that if 
dark matter is composed of scalars or Majorana fermions, annihilation into 
a neutrino-antineutrino pair is highly disfavored from angular momentum 
considerations. For all final states except neutrinos, these bounds are 
comparable to or stronger than the corresponding limits from IceCube.  
The current generation of direct detection 
experiments~\cite{Aprile:2009zzb}, ~\cite{Fiorucci:2009ak} have the 
potential to strengthen this bound by an additional two orders of 
magnitude. In such a scenario the maximum number of events will be reduced 
well below the background from atmospheric neutrinos, and the 
spin-independent case will go out of reach of current generation neutrino 
telescopes. We also see from the figures that the corresponding bounds on 
the event rates for models with purely spin-dependent interactions are 
much weaker, and are not competitive at all with the current limits from 
neutrino telescopes.

\begin{figure}[htp]
  \begin{center}
    \includegraphics[width=0.8 \textwidth]{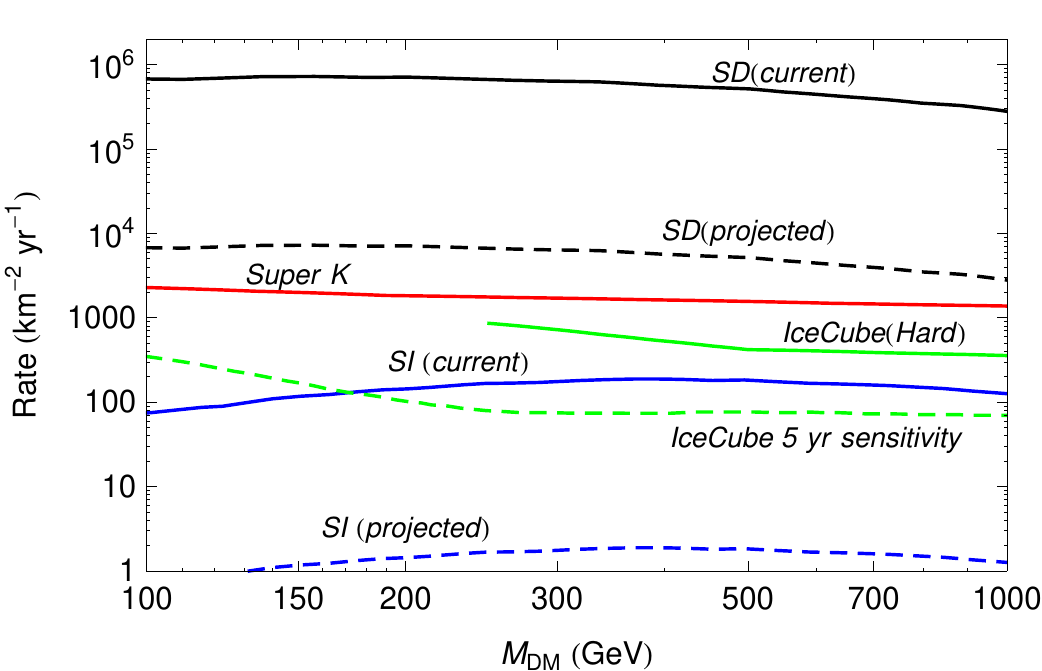}
  \end{center}
  \caption{ Shown are bounds and projected bounds on the event rate from 
  Super Kamiokande
  and IceCube. Also plotted are limits on the event rate obtained from 
  null
  results in direct detection experiments, both at present and 
  projected
  (two orders of magnitude stronger). Solid lines denote current
  bounds and dashed lines denote projections. The final state was
  assumed to be $W^+ W^-$ and the threshold energy set at 1 GeV.}
  \label{fig:nubounds}
\end{figure}

In Fig \ref{fig:nubounds} we plot the current IceCube \cite{Abbasi:2009uz} 
and Super-Kamiokande \cite{Desai:2007ra} bounds on the event rates 
assuming annihilation to the $W^+ W^-$ final state, together with bounds 
on the event rate from direct detection experiments. It is clear from 
these plots that the direct detection bounds on spin-independent 
interactions lead to a tighter constraint on event rates than the current 
IceCube limit. With the operation of newer detectors, the spin-independent 
bound will only become stronger, and consequently is expected to be beyond 
the reach of IceCube 5 year sensitivity \cite{clercq2008}. The 
spin-dependent case, on the other hand, is seen to allow a much higher 
flux. In fact, by far the strongest bound on the spin-dependent 
interaction (assuming $W^+W^-$ annihilation) is seen to come from IceCube.

If the number of events observed in neutrino telescopes were to exceed the 
upper bounds corresponding to spin-independent interactions, the 
implication is that the cross section of dark matter with nucleons is 
dominated by spin-dependent interactions. In the next section, we consider 
the dark matter models that have this property. However, before we do so, 
we briefly consider the effect of relaxing some of the assumptions that 
went into obtaining these bounds.
\begin{itemize}
\item
We restricted ourselves to WIMP annihilations into two particle SM final 
states. Relaxing this assumption by allowing multi-body 
SM final states does not affect the upper bound. The reason is that for 
any given WIMP mass, the neutrinos arising from the decays of a 
multi-particle final state each carry correspondingly less energy. Since 
the neutrino-nucleus cross section scales as the energy of the incident 
neutrino, as does the range of the muon produced in the collision, these 
effects reduce the signal more than enough to compensate for the increased 
number of neutrinos.
\item
We did not consider the possibility of annihilations into non-SM final 
states composed of new particles yet to be discovered. These could 
subsequently decay into SM states and eventually into neutrinos, thereby 
generating a signal. However, we do not expect this to alter the upper 
bound, since each SM particle will carry less energy than if annihilation 
had occurred directly to two SM particles, with a corresponding reduction 
in the signal.
\item
  We ignored effects like Sommerfeld enhancement
  \cite{sommerfeld1931,Cirelli:2007xd,ArkaniHamed:2008qn,Bedaque:2009ri}, 
  which can
  significantly affect the annihilation rate.  This is justified
  because the upper bound on the signal is controlled only by the
  capture rate, and does not depend on the annihilation rate.
\item
  We assumed a specific value for the local halo density. Our bounds
  on the event rates are insensitive to this value since it affects
  the capture rate and the direct detection bounds in exactly the same
  way.
\item
  We assumed a specific distribution of dark matter velocities. The
  result is in fact sensitive to this assumption since it affects the
  capture rate and the direct detection bounds somewhat differently.
  However, we do not expect the uncertainties in this to affect our
  conclusions significantly.
\item
  We restricted our considerations to elastic dark matter. Our result
  does not apply to inelastic dark
  matter~\cite{TuckerSmith:2001hy,TuckerSmith:2004jv,Cui:2009xq} or
  more generally to any type of form-factor dark
  matter~\cite{Feldstein:2009tr, Chang:2009yt, Bai:2009cd} since the
  relationship between the capture rate and the direct detection
  bounds is now modified~\cite{Nussinov:2009ft,Menon:2009qj}.
\item
Finally, note that in obtaining these bounds we have not restricted 
ourselves to thermal dark matter, but have also allowed for the 
possibility that dark matter is composed of non-thermal WIMPs. 
\end{itemize}
\begin{table}
  \begin{minipage}[htp]{\textwidth}
    \renewcommand{\thempfootnote}{\fnsymbol{mpfootnote}}
    \begin{center}
      \begin{tabular}{ |>{\centering}m{1.4in}| >{\centering}m{0.6in}|
	>{\centering}m{1.5in}|c|}
	\hline
	Dark Matter&Mediator & Process & Scattering\\
	\hline
	\multirow{3}{*}{Scalar}
	&$Z,Z'$& 
	\rule{0pt}{25pt}
	\begin{fmffile}{tab-s1}
	  \begin{fmfgraph*}(12,8)
	    \fmfleftn{ia}{2} \fmfrightn{oa}{2}
	    \fmf{dashes}{ia1,va1,ia2}
	    \fmf{boson}{va1,va2}
	    \fmf{plain}{oa1,va2,oa2}
	  \end{fmfgraph*}
	\end{fmffile}
	&
	SI\\
	\cline{2-4}
	&$h$
	&
	\rule{0pt}{25pt}
	\begin{fmffile}{tab-s2}
	  \begin{fmfgraph*}(12,8)
	    \fmfleftn{ib}{2} \fmfrightn{ob}{2}
	    \fmf{dashes}{ib1,vb1,ib2}
	    \fmf{dashes}{vb1,vb2}
	    \fmf{plain}{ob1,vb2,ob2}
	  \end{fmfgraph*}
	\end{fmffile}
	&  SI\\
	\cline{2-4}
	&$Q$&
	\rule{0pt}{25pt}
	\begin{fmffile}{tab-s3}
	  \begin{fmfgraph*}(12,8)
	    \fmfleftn{id}{2} \fmfrightn{od}{2}
	    \fmf{phantom}{id1,vd1}
	    \fmf{phantom}{id2,vd2}
	    \fmf{dashes,tension=0}{id2,vd1}
	    \fmf{dashes,tension=0}{id1,vd2}
	    \fmf{plain}{vd1,vd2}
	    \fmf{plain}{od1,vd1}
	    \fmf{plain}{vd2,od2}
	  \end{fmfgraph*}
	  ,
	  \begin{fmfgraph*}(12,8)
	    \fmfleftn{id}{2} \fmfrightn{od}{2}
	    \fmf{dashes}{id1,vd1}
	    \fmf{dashes}{id2,vd2}
	    \fmf{plain}{vd1,vd2}
	    \fmf{plain}{od1,vd1}
	    \fmf{plain}{vd2,od2}
	  \end{fmfgraph*}
	\end{fmffile}
	&  SI\\
	\hline
	\multirow{4}{*}{Dirac Fermion}
	&$Z$,$Z'$& 
	\rule{0pt}{25pt}
	\begin{fmffile}{tab-f1}
	  \begin{fmfgraph*}(12,8)
	    \fmfleftn{ia}{2} \fmfrightn{oa}{2}
	    \fmf{plain}{ia1,va1,ia2}
	    \fmf{boson}{va1,va2}
	    \fmf{plain}{oa1,va2,oa2}
	  \end{fmfgraph*}
	\end{fmffile}
	&  SI, SD\footnote[2]
	{Can be primarily SD for specific choices of $Z'$
	charges}
	\\
	\cline{2-4}
	&$h$
	&
	\rule{0pt}{25pt}
	\begin{fmffile}{tab-f2}
	  \begin{fmfgraph*}(12,8)
	    \fmfleftn{ib}{2} \fmfrightn{ob}{2}
	    \fmf{plain}{ib1,vb1,ib2}
	    \fmf{dashes}{vb1,vb2}
	    \fmf{plain}{ob1,vb2,ob2}
	  \end{fmfgraph*}
	\end{fmffile}
	&  SI\\
	\cline{2-4}
	&$X$&
	\rule{0pt}{25pt}
	\begin{fmffile}{tab-f3}
	  \begin{fmfgraph*}(12,8)
	    \fmfleftn{ic}{2} \fmfrightn{oc}{2}
	    \fmf{plain}{ic1,vc1}
	    \fmf{plain}{vc2,ic2}
	    \fmf{boson,tension=0.5}{vc1,vc2}
	    \fmf{phantom}{oc1,vc1}
	    \fmf{phantom}{vc2,oc2}
	    \fmf{plain,tension=0}{oc1,vc2}
	    \fmf{plain,tension=0}{vc1,oc2}
	  \end{fmfgraph*}
	  ,
	  \begin{fmfgraph*}(12,8)
	    \fmfleftn{id}{2} \fmfrightn{od}{2}
	    \fmf{plain}{id1,vd1}
	    \fmf{plain}{id2,vd2}
	    \fmf{boson,tension=0.5}{vd1,vd2}
	    \fmf{plain}{od1,vd1}
	    \fmf{plain}{vd2,od2}
	  \end{fmfgraph*}
	\end{fmffile}
	&  SI, SD\\
	\cline{2-4}
	&$\Phi$&
	\rule{0pt}{25pt}
	\begin{fmffile}{tab-f4}
	  \begin{fmfgraph*}(12,8)
	    \fmfleftn{id}{2} \fmfrightn{od}{2}
	    \fmf{plain}{id1,vd1}
	    \fmf{plain}{vd2,id2}
	    \fmf{dashes}{vd1,vd2}
	    \fmf{phantom}{od1,vd1}
	    \fmf{phantom}{vd2,od2}
	    \fmf{plain,tension=0}{od1,vd2}
	    \fmf{plain,tension=0}{vd1,od2}
	  \end{fmfgraph*}
	  ,
	  \begin{fmfgraph*}(12,8)
	    \fmfleftn{id}{2} \fmfrightn{od}{2}
	    \fmf{plain}{id1,vd1}
	    \fmf{plain}{id2,vd2}
	    \fmf{dashes}{vd1,vd2}
	    \fmf{plain}{od1,vd1}
	    \fmf{plain}{vd2,od2}
	  \end{fmfgraph*}
	\end{fmffile}
	&  SI, SD\\
	\hline
	\multirow{4}{*}{Majorana Fermion}
	&$Z$,$Z'$& 
	\rule{0pt}{25pt}
	\begin{fmffile}{tab-fm1}
	  \begin{fmfgraph*}(12,8)
	    \fmfleftn{ia}{2} \fmfrightn{oa}{2}
	    \fmf{plain}{ia1,va1,ia2}
	    \fmf{boson}{va1,va2}
	    \fmf{plain}{oa1,va2,oa2}
	  \end{fmfgraph*}
	\end{fmffile}
	&  SD\\
	\cline{2-4}
	&$h$
	&
	\rule{0pt}{25pt}
	\begin{fmffile}{tab-fm2}
	  \begin{fmfgraph*}(12,8)
	    \fmfleftn{ib}{2} \fmfrightn{ob}{2}
	    \fmf{plain}{ib1,vb1,ib2}
	    \fmf{dashes}{vb1,vb2}
	    \fmf{plain}{ob1,vb2,ob2}
	  \end{fmfgraph*}
	\end{fmffile}
	&  SI\\
	\cline{2-4}
	&$X$&
	\begin{fmffile}{tab-fm3}
	  \rule{0pt}{19pt}
	  \begin{tabular}{m{29pt}m{0pt}m{29pt}}
	    \begin{fmfgraph*}(12,8)
	      \fmfleftn{ic}{2} \fmfrightn{oc}{2}
	      \fmf{plain}{ic1,vc1}
	      \fmf{plain}{vc2,ic2}
	      \fmf{boson,tension=0.5}{vc1,vc2}
	      \fmf{phantom}{oc1,vc1}
	      \fmf{phantom}{vc2,oc2}
	      \fmf{plain,tension=0}{oc1,vc2}
	      \fmf{plain,tension=0}{vc1,oc2}
	    \end{fmfgraph*}
	    &+&
	    \begin{fmfgraph*}(12,8)
	      \fmfleftn{ic}{2} \fmfrightn{oc}{2}
	      \fmf{boson,tension=0.5}{vc1,vc2}
	      \fmf{plain}{ic1,vc1}
	      \fmf{plain}{vc2,ic2}
	      \fmf{plain}{oc1,vc1}
	      \fmf{plain}{vc2,oc2}
	    \end{fmfgraph*}
	  \end{tabular}
	\end{fmffile}
	& SD in chiral limit\\
	\cline{2-4}
	&$\Phi$&
	\begin{fmffile}{tab-fm4}
	  \rule{0pt}{19pt}
	  \begin{tabular}{m{29pt}m{0pt}m{29pt}}
	    \begin{fmfgraph*}(12,8)
	      \fmfleftn{id}{2} \fmfrightn{od}{2}
	      \fmf{plain}{id1,vd1}
	      \fmf{plain}{vd2,id2}
	      \fmf{dashes}{vd1,vd2}
	      \fmf{phantom}{od1,vd1}
	      \fmf{phantom}{vd2,od2}
	      \fmf{plain,tension=0}{od1,vd2}
	      \fmf{plain,tension=0}{vd1,od2}
	    \end{fmfgraph*}
	    &+&
	    \begin{fmfgraph*}(12,8)
	      \fmfleftn{id}{2} \fmfrightn{od}{2}
	      \fmf{plain}{vd1,id1}
	      \fmf{plain}{id2,vd2}
	      \fmf{dashes}{vd1,vd2}
	      \fmf{plain}{od1,vd1}
	      \fmf{plain}{vd2,od2}
	    \end{fmfgraph*}
	  \end{tabular}
	\end{fmffile}
	&  SD  in chiral limit\\
	\hline
	\multirow{2}{*}{Real Vector}
	&$h$
	&
	\rule{0pt}{25pt}
	\begin{fmffile}{tab-v1}
	  \begin{fmfgraph*}(12,8)
	    \fmfleftn{ib}{2} \fmfrightn{ob}{2}
	    \fmf{boson}{ib1,vb1,ib2}
	    \fmf{dashes}{vb1,vb2}
	    \fmf{plain}{ob1,vb2,ob2}
	  \end{fmfgraph*}
	\end{fmffile}
	&  SI\\
	\cline{2-4}
	&$Q$&
	\begin{fmffile}{tab-v2}
	  \rule{0pt}{19pt}
	  \begin{tabular}{m{22pt}m{0pt}m{22pt}}
	    \begin{fmfgraph*}(12,8)
	      \fmfleftn{ic}{2} \fmfrightn{oc}{2}
	      \fmf{boson}{ic1,vc1}
	      \fmf{boson}{ic2,vc2}
	      \fmf{plain}{vc1,vc2}
	      \fmf{plain}{oc1,vc1}
	      \fmf{plain}{vc2,oc2}
	    \end{fmfgraph*}
	    &+&
	    \begin{fmfgraph*}(12,8)
	      \fmfleftn{id}{2} \fmfrightn{od}{2}
	      \fmf{boson}{id1,vd1}
	      \fmf{boson}{vd2,id2}
	      \fmf{plain}{vd2,vd1}
	      \fmf{phantom}{od1,vd1}
	      \fmf{phantom}{vd2,od2}
	      \fmf{plain,tension=0}{od1,vd2}
	      \fmf{plain,tension=0}{vd1,od2}
	    \end{fmfgraph*}
	  \end{tabular}
	\end{fmffile}
	&  SD in chiral limit\\
	\hline
	\multirow{3}{*}{Complex Vector}
	&$Z,Z'$& 
	\rule{0pt}{25pt}
	\begin{fmffile}{tab-v1b}
	  \begin{fmfgraph*}(12,8)
	    \fmfleftn{ia}{2} \fmfrightn{oa}{2}
	    \fmf{boson}{ia1,va1,ia2}
	    \fmf{boson}{va1,va2}
	    \fmf{plain}{oa1,va2,oa2}
	  \end{fmfgraph*}
	\end{fmffile}
	&
	SI\\
	\cline{2-4}
	&$h$
	&
	\rule{0pt}{25pt}
	\begin{fmffile}{tab-v2b}
	  \begin{fmfgraph*}(12,8)
	    \fmfleftn{ib}{2} \fmfrightn{ob}{2}
	    \fmf{boson}{ib1,vb1,ib2}
	    \fmf{dashes}{vb1,vb2}
	    \fmf{plain}{ob1,vb2,ob2}
	  \end{fmfgraph*}
	\end{fmffile}
	&  SI\\
	\cline{2-4}
	&$Q$&
	\rule{0pt}{25pt}
	\begin{fmffile}{tab-v3b}
	  \begin{fmfgraph*}(12,8)
	    \fmfleftn{id}{2} \fmfrightn{od}{2}
	    \fmf{boson}{id1,vd1}
	    \fmf{boson}{id2,vd2}
	    \fmf{plain}{vd1,vd2}
	    \fmf{plain}{od1,vd1}
	    \fmf{plain}{vd2,od2}
	  \end{fmfgraph*}
	  ,
	  \begin{fmfgraph*}(12,8)
	    \fmfleftn{id}{2} \fmfrightn{od}{2}
	    \fmf{boson}{id1,vd1}
	    \fmf{boson}{vd2,id2}
	    \fmf{plain}{vd2,vd1}
	    \fmf{phantom}{od1,vd1}
	    \fmf{phantom}{vd2,od2}
	    \fmf{plain,tension=0}{od1,vd2}
	    \fmf{plain,tension=0}{vd1,od2}
	  \end{fmfgraph*}
	\end{fmffile}
	&  SI, SD\\
	\hline
      \end{tabular}
    \end{center}
  \caption{ A summary of results for WIMP-nucleon scattering, for each
  dark matter candidate and mediator~\cite{Agrawal:2010fh}. In the
  Feynman diagrams, scalars are represented by dashed lines, fermions
  by solid lines and vector bosons by wavy lines.  Of the mediators,
  $h$, $Z'$ and the SM $Z$ are neutral under both electromagnetism and
  color, while $X$, $\Phi$ and $Q$ transform as triplets under color
  and carry electric charge.}
  \label{tab:spind-summary}
  \end{minipage}
\end{table}
\afterpage{\clearpage}
\section{Collider Implications}

In this section we consider the implications for the LHC if the 
WIMP-nucleon spin-dependent cross-section lies within the parameter space 
that will be probed by IceCube. The dark matter candidates that can 
naturally give rise to primarily spin-dependent interactions in a 
renormalizable theory have been classified~\cite{Agrawal:2010fh}, and the 
results are shown in Table \ref{tab:spind-summary}. From the table it is 
clear that if the interactions of WIMPs with nucleons are primarily 
spin-dependent, the natural dark matter candidates are either Majorana 
fermions or real vector bosons. Furthermore, corresponding to each dark matter 
candidate, the spins and quantum numbers of the particles mediating the 
tree-level spin-dependent WIMP-nucleon interaction are also constrained. 
In this section we study the range of masses of the dark matter particle, 
and also of the mediating particle, that lead to an observable signal from 
spin-dependent interactions at IceCube. We will establish that these 
particles naturally tend to lie within the energy range that will be 
probed by the LHC. Our approach will be to study the various candidate 
theories in turn, and study the range of particle masses in each case.

\subsection{Majorana Fermion}
The effective operator that gives rise to spin-dependent scattering in
this case can be shown to be
\begin{align}
  L_{\text{eff}} &= 
  d_q\, \bar{\chi} 
  \gamma^\mu\gamma^5
  \chi\;
  \bar{q}
  \gamma_\mu\gamma^5
  q
  \label{eq:eff-op}
  ,
\end{align}
where $\chi$ is the Majorana spinor corresponding to the dark matter 
particle and $q$ represents a quark field.
The corresponding WIMP-nucleus cross-section is given by
\begin{align}
  \sigma_0
  =
  \frac{16 m_{\chi}^2m_N^2}{\pi(m_{\chi}+m_N)^2}
  \left[\sum_{q=u,d,s}{d_q\lambda_q}\right]^2
  J_N(J_N+1)
  \label{eq:maj-cs}
  .
\end{align}
For scattering off a free nucleon $\lambda_q = \Delta_q^n$. The
quantities $\Delta_q$ are the spin-fraction of the nucleon spin
carried by quark $q$, listed in Table \ref{tab:qspin}. $J_N$ is the
angular momentum of the nucleus, and is equal to $\frac12$ for free
nucleons.

\begin{table}
  \begin{center}
  \begin{tabular}{c|cc}
    & proton & neutron\\
    \hline
    $\Delta_u$ &$  0.78\pm0.02$ & $-0.48\pm0.02$\\
    $\Delta_d$ &$ -0.48\pm0.02$ & $ 0.78\pm0.02$\\
    $\Delta_s$ &$ -0.15\pm0.02$ & $-0.15\pm0.02$\\
  \end{tabular}
  \end{center}
  \caption{Quark spin fraction in the proton and neutron~
  \cite{Ellis:2000ds,Mallot:1999qb}}
  \label{tab:qspin}
\end{table}

We see from the table that if dark matter is composed of Majorana 
fermions, 
spin-dependent WIMP-nucleon scattering can arise through any of
\begin{itemize}
\item
t-channel vector exchange,
\item
s- and u-channel vector exchange, or
\item
s- and u-channel scalar exchange
\end{itemize}
If the process occurs via the s or u-channels the mediating particle 
necessarily carries SM color. For a t-channel process the mediating vector 
boson may either be the SM $Z$, or a new $Z'$. If it is the SM $Z$, the 
dark matter particle carries weak charge. Therefore we see that each of 
these processes is associated either with a new particle charged under the 
SM gauge groups, or with a new $Z'$ gauge boson.

The values of $d_u$, $d_d$ and $d_s$ which appear in equation 
\eqref{eq:maj-cs} depend on the flavor structure of the underlying theory, 
and must be consistent with constraints on flavor-violating processes. 
Dark matter scattering mediated by a $Z'$ will automatically satisfy these 
constraints provided that the couplings of the $Z'$ are flavor diagonal. 
This implies that in these models $d_d = d_s$. For theories where 
WIMP-nucleon scattering is mediated by $X_{\mu}$ or $\Phi$, the flavor 
structure is more complicated. In general, a GIM mechanism which 
incorporates either multiple flavors of the dark matter particle or 
multiple flavors of the mediators is required in order to ensure that 
flavor bounds are satisfied. For concreteness, in what follows we assume 
that there are multiple mediators. Flavor bounds are then satisfied 
provided that the mediators corresponding to different flavors are 
quasi-degenerate and their couplings are flavor-diagonal.

\subsubsection* {t-channel vector exchange}
\begin{figure}[h]
  \begin{center}
    \begin{fmffile}{ma}
      \begin{fmfgraph*}(30,20)
	\fmfleftn{ia}{2} \fmfrightn{oa}{2}
	\fmflabel{$\chi$}{ia1}
	\fmflabel{$\chi$}{ia2}
	\fmflabel{$q$}{oa1}
	\fmflabel{$q$}{oa2}
	%\fmflabel{$e_-$}{i1}\fmflabel{$e_+$}{i2}
	%\fmflabel{$\mu_+$}{o1}
	%\fmflabel{$\nu_{\mu}$}{o2}
	\fmf{fermion, label=$p_1$}{ia1,va1}
	\fmf{fermion, label=$p_3$}{va1,ia2}
	\fmf{boson, label=$Z$}{va1,va2}
	\fmf{fermion,label=$p_2$}{oa1,va2}
	\fmf{fermion,label=$p_4$}{va2,oa2}
      \end{fmfgraph*}
    \end{fmffile}
  \end{center}
  \caption{Majorana dark matter scattering through t-channel vector 
exchange}
  \label{fig:ma}
\end{figure}
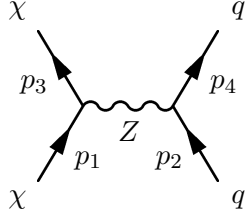

We begin with t-channel exchange of a vector. The Lagrangian corresponding 
to this process can be written in four-component language as,
\begin{align}
  \mathcal{L}&=
  -\frac{1}{4}\mathcal{F}^{\mu\nu}\mathcal{F}_{\mu\nu}
  +\frac{1}{2}m_Z^2\;Z^{\mu}Z_{\mu}
  +\bar{\chi}\gamma^{\mu}(-\beta\gamma^5)\chi Z_{\mu}
  +\bar{q}\gamma^{\mu}(\widetilde{\alpha}-\widetilde\beta\gamma^5)qZ_{\mu}
  .
\end{align}
For simplicity, in this expression flavor indices have been suppressed. 
It is straightforward to verify that the coefficient of the effective operator 
defined in equation \eqref{eq:eff-op} is
\begin{align}
  d_q &= -\frac{\beta \widetilde{\beta_q}}{m_{Z}^2}
  .
\end{align}
Here $Z$ could represent either the SM $Z$ or a new $Z'$ under which the 
SM matter fields are charged. 
For the Standard Model $Z$,
\begin{align}
  d_u = -d_d = -d_s
  .
\end{align}
We choose to assign the $Z'$ charges in the same way, because for a given 
WIMP-proton cross-section this choice of signs corresponds to the highest 
value of the mass of the exchanged particle, and is therefore the 
conservative assumption. It is also naturally consistent with flavor 
bounds on new physics, as noted above. The physics of the two cases is 
however very different, and so we consider them separately.

If the WIMP carries charge under the $Z$, it must either constitute the 
neutral component of a single SM SU(2$)_{\rm L}$ representation or arise 
as a linear combination of the neutral components of different 
representations of SU(2$)_{\rm L}$. Majorana fermion dark matter that 
arises from a single SU(2$)_{\rm L}$ representation is very tightly 
constrained by direct detection experiments. For example, Majorana 
neutrino dark matter has been excluded by Xenon in the mass range 
considered here \cite{Angle:2008we}. We will therefore not consider this 
possibility further.

Although theories where the dark matter arises as a linear combination of 
the neutral components of different SU(2$)_{\rm L}$ representations are 
also constrained by direct detection experiments, the bounds are 
significantly weaker. In general there are both spin-dependent 
WIMP-nucleon interactions arising from $Z$ exchange as well as 
spin-independent interactions arising from Higgs exchange. The magnitudes 
of these two different contributions to the WIMP-nucleon cross section are 
not in general independent, but are correlated~\cite{Cohen:2010gj}.

\begin{figure}[htp]
  \begin{center}
    \includegraphics[scale=1.0]{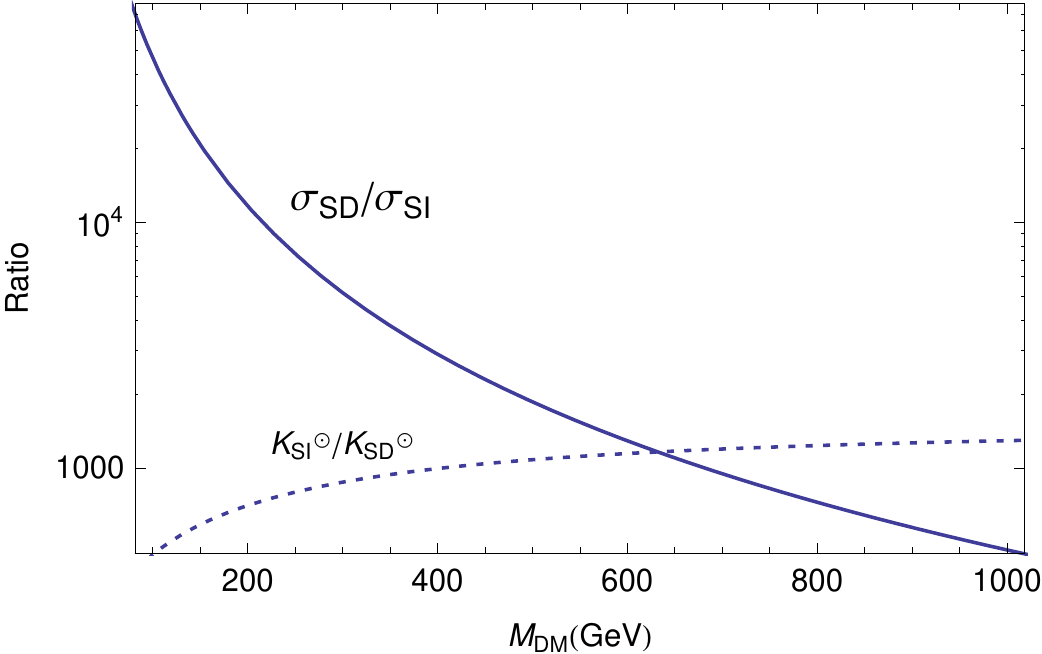}
  \end{center}
  \caption{The ratio of spin-dependent and spin-independent
  cross-sections in the benchmark model for WIMP-nucleon
  scattering through the SM $Z$ boson. We also show the ratio of
  capture efficiencies, $K^\odot_{SI}/K^\odot_{SD}$. To the left of
  the intersection, the neutrinos from SD interactions dominate.}
  \label{fig:sisdrat}
\end{figure}

To understand this correlation we work in the benchmark model of 
reference~\cite{Agrawal:2010fh}. This model consists of two SU(2$)_{\rm 
L}$ fermion doublets $\xi$ and $\xi^c$ which have equal and opposite 
hypercharges ${\rm Y} = \pm \frac12$. The dark matter particle arises as a 
linear combination of the neutral components of $\xi$ and $\xi^c$. The 
mass matrix consists of a Dirac term linking $\xi$ and $\xi^c$, together 
with a Majorana term that arises from a non-renormalizable interaction of 
the form $ (H^\dagger \xi)^2$. When the SM Higgs doublet $H$ gets a VEV, 
the neutral component of $\xi$ acquires a Majorana mass. It also obtains a 
Yukawa coupling to the physical Higgs field $h$. Such a non-renormalizable 
operator may be generated by integrating out a SM singlet. The Lagrangian 
for the neutral components of $\xi$ and $\xi^c$ takes the form
\begin{align}
  \mathcal{L} \supset 
  -
  \frac{g}{\cos\theta_W}
  \left(\bar{\xi}_0 \frac12 \bar{\sigma}^\mu \xi_0
  -\bar{\xi}_0^c \frac12 \bar{\sigma}^\mu \xi_0^c\right)
  Z_\mu
  -
  \left[
  \frac12
  \left(
  \begin{array}{cc}
    \xi^c_0 & \xi_{0}
  \end{array}
  \right)
  \left(
  \begin{array}{cc}
    0 & M\\
    M & m
  \end{array}
  \right)
  \left(
  \begin{array}{c}
    \xi^c_0\\
    \xi_{0}
  \end{array}
  \right)
  + y_\xi
  \xi_0 \xi_0 h
  +{\rm h.c.}
  \right] 
  .
\end{align}
Here $y_\xi=m/v$ where $v = 246$ GeV is the electroweak VEV, and $h$ is 
the physical Higgs field of the SM. The lighter of the two mass 
eigenstates $\xi_D$ is the dark matter particle.

\begin{align}
  \xi_D = 
  \cos \phi\; \xi_0 + \sin\phi\; \xi_0^c 
  ,
\end{align}
where $\phi$ is the mixing angle. 
The couplings to the $Z$ and to the SM Higgs in the mass basis take the 
form,
\begin{align}
  \mathcal{L} \supset 
  -\frac{g\cos2\phi}{2\cos\theta_W}
  \bar{\xi}_D\bar{\sigma}^\mu  \xi_D
  Z_\mu 
  -
  \left[
  y_\xi \cos^2\phi\,
  \xi_D \xi_D h
  + {\rm h.c.}
  \right]
  .
\end{align} 
In terms of a four-component Majorana fermion $\chi$, the Lagrangian 
becomes
\begin{align}
  \mathcal{L} \supset 
  \frac{g\cos2\phi}{4\cos\theta_W}
  \bar{\chi}\gamma^\mu \gamma^5\chi
  Z_\mu 
  - y_\xi \cos^2 \phi\,
  \bar{\chi} \chi h
  .
\end{align} 

In order to understand the correlation between the spin-dependent and 
spin-independent contributions to the WIMP-nucleon cross section let us estimate 
the mixing in the limit $m\ll M$. The coupling to the $Z$ is suppressed by $\cos 
2\phi$, which in this limit is simply $m/2M$, while the dark matter mass is 
approximately equal to $M$.  On the other hand the coupling to the Higgs is 
proportional to $m/v$, where $v = 246$ GeV is the electroweak VEV. Therefore in 
this limit the ratio of the spin-dependent WIMP-nucleon cross section to the 
corresponding spin-independent cross section is fixed for any given value of $M$. 
We have plotted the ratio of these cross sections as a function of the WIMP mass 
in Fig \ref{fig:sisdrat}, in the limit that the spin-independent cross section is 
at its current bound and assuming a Higgs mass of 120 GeV. Note that the 
spin-dependent cross section never exceeds the spin-independent cross section by 
more than a factor of about $10^5$ for dark matter masses greater than 150 GeV. We 
have also shown on the same plot the ratio of capture efficiencies in the 
spin-independent and spin-dependent cases. Clearly, spin-independent capture 
dominates for WIMP masses larger than about $600$ GeV.  Therefore a limit on the 
spin-independent WIMP-nucleon cross section directly translates into an upper 
bound on the event rate in neutrino telescopes in this benchmark model. We expect 
that the same conclusion will apply to the more general class of models where 
spin-dependent scattering is mediated by the SM $Z$.

\begin{figure}[h]
  \begin{center}
    \includegraphics[scale=1.0]{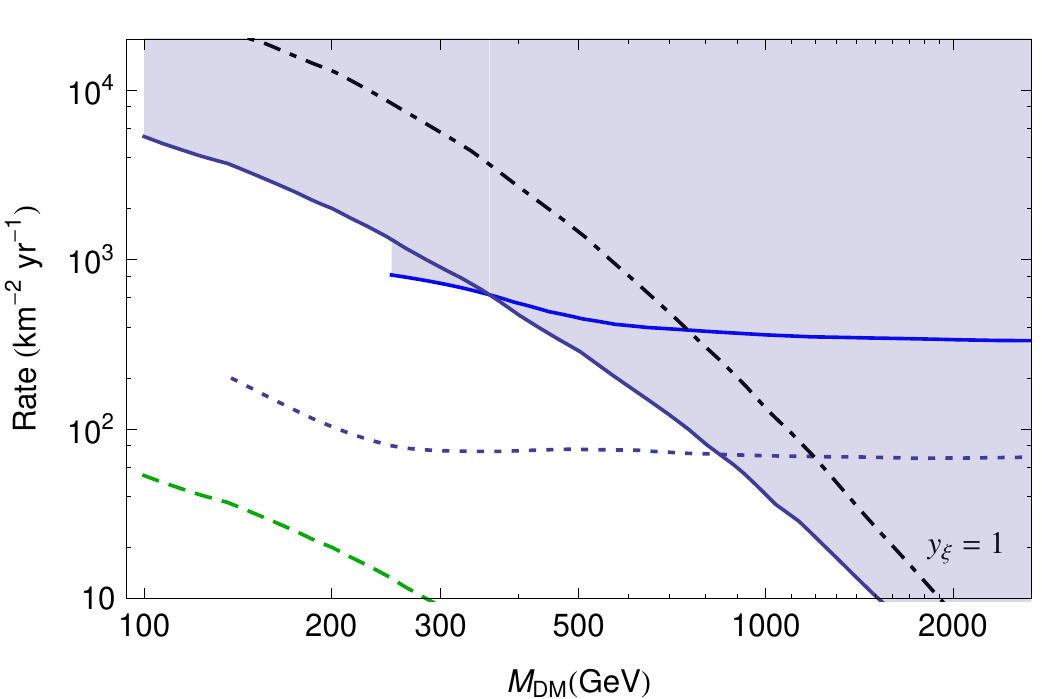}
  \end{center}
  \caption{Estimates for the rates in IceCube for a Majorana dark
  matter coupling through the Standard Model $Z$, annihilating into
  $W^+W^-$ ($E_{th}=10$ GeV). The coupling was assumed to be at the CDMS
  spin-independent bound (blue solid line) and at a projected
  spin-independent bound (green dashed line). The IceCube current
  bound and 5 year sensitivity (dotted line) are also shown. The
  dot-dashed line corresponds to the WIMP-Higgs Yukawa coupling
  $y_\xi=1$. The shaded region has been excluded by either direct
  detection experiments or IceCube.}
  \label{fig:icezsd}
\end{figure}

In Fig \ref{fig:icezsd} we have plotted the current upper limit on the 
event rate from direct detection experiments in this scenario as a 
function of the dark matter mass, assuming annihilation is entirely to the 
$W^+W^-$ final state. We have also shown the current and future 
experimental bounds from IceCube. Note that in this and subsequent plots 
we do not require that annihilation and capture are mediated by the same 
particle, but allow for the possibility that these are mediated by 
different particles. In particular, dark matter annihilation exclusively 
to $W^+ W^-$ is not possible if mediated by the SM $Z$. We see that a 
signal close to the present IceCube bound implies a dark matter mass less 
than about 400 GeV, which is promising for the LHC. More generally, in 
most of the parameter space that is within the 5-year sensitivity of 
IceCube, the dark matter particle and its charged partners $\xi^+$ and 
${\xi^c}^-$ have masses below ~1 TeV. The interesting region of parameter 
space lies below the line corresponding to $y_{\xi} = 1$, indicating that 
it is under perturbative control. In the same figure, we have also shown 
the effect on the event rate if the direct detection bounds improve by two 
orders of magnitude, as expected. In such a scenario, the signal in this 
class of theories would fall below the sensitivity of IceCube. Although 
this analysis has been performed for a specific benchmark model, we expect 
that similar conclusions hold in general for spin-dependent interactions 
mediated by the SM $Z$.

We now turn to the case of the $Z'$. There is now much more freedom with 
regard to charge assignments and the overall strength of the interaction. 
Also, there need not be a spin-independent contribution to the cross 
section from scalar exchange.  However, experimental constraints on the 
masses and couplings of new $Z'$ gauge bosons disfavor large event rates. 
As in the case of the SM $Z$, we set $d_u=-d_d=-d_s$, since this choice is 
naturally consistent with flavor constraints on new physics, and yields a 
conservative estimate for the mediator mass.  The values $\beta$ and 
$\widetilde{\beta}$ equal to a $\frac12$ correspond to chiral fermions 
having unit charge under the $Z'$.  We use these as representative values. 
In Fig \ref{fig:lhcu} we have shown the range of values of the $Z'$ 
mass that would lead to a signal in neutrino telescopes at the current 
bound, or at the 5-year sensitivity.  Unfortunately, a large portion of 
the allowed range of masses is disfavored by precision electroweak 
measurements and by bounds on direct production and four-point 
interactions~\cite{Appelquist:2002mw,Amsler:2008zzb}, except perhaps for 
very specific $Z'$ charge assignments.

\subsubsection* {s- and u-channel vector exchange}
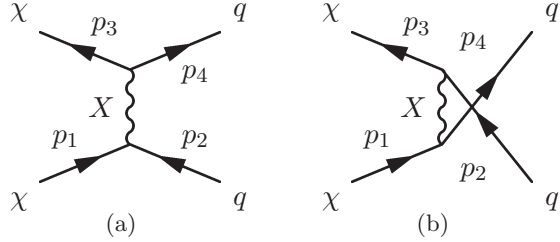
\begin{figure}[h]
  \begin{center}
    \begin{fmffile}{mb}
      \subfloat[]{
      \label{fig:mba}
      \begin{fmfgraph*}(30,20)
	\fmfleftn{ix}{2} \fmfrightn{ox}{2}
	\fmflabel{$\chi$}{ix1}
	\fmflabel{$\chi$}{ix2}
	\fmflabel{$q$}{ox1}
	\fmflabel{$q$}{ox2}
	\fmf{fermion, label=$p_1$}{ix1,vx1}
	\fmf{fermion, label=$p_3$}{vx2,ix2}
	\fmf{boson, label=$X$}{vx1,vx2}
	\fmf{fermion, label=$p_2$}{ox1,vx1}
	\fmf{fermion, label=$p_4$}{vx2,ox2}
      \end{fmfgraph*}}
      \hspace{0.3in}
      \subfloat[]{
      \label{fig:mbb}
      \begin{fmfgraph*}(30,20)
	\fmfleftn{iy}{2} \fmfrightn{oy}{2}
	\fmflabel{$\chi$}{iy1}
	\fmflabel{$\chi$}{iy2}
	\fmflabel{$q$}{oy1}
	\fmflabel{$q$}{oy2}
	\fmf{fermion, label=$p_1$}{iy1,vy1}
	\fmf{fermion, label=$p_3$}{vy2,iy2}
	\fmf{boson, label=$X$}{vy1,vy2}
	\fmf{phantom,label=$p_2$,l.d=0.01mm,l.s=left}{oy1,vy1}
	\fmf{phantom,label=$p_4$,l.d=0.01mm,l.s=left}{vy2,oy2}
	\fmf{fermion,tension=0}{oy1,vy2}
	\fmf{fermion,tension=0}{vy1,oy2}
      \end{fmfgraph*}}
    \end{fmffile}
  \end{center}
  \caption{Majorana dark matter scattering through s- and u-channel vector 
           exchange}
  \label{fig:mb}
\end{figure}
We now move on to s- and u-channel exchange of a colored vector
particle $X$.
The Lagrangian corresponding to this process is
\begin{align}
  \mathcal{L}
  &=
  -\frac{1}{2}\left|\partial_\mu X_\nu - \partial_\nu X_\mu\right|^2
  +m_X^2\;X^\dagger_\mu X^\mu
  +\bar{\chi}\gamma^{\mu}(\alpha-\beta\gamma^5)q
  X_{\mu}
  +\bar{q}\gamma^{\mu}(\alpha^*-\beta^*\gamma^5)\chi
  X_{\mu}^\dagger
  .
\end{align}

The co-efficient of the effective operator in this case is given by
\begin{align}
  d_q &= 
  -\frac{|\alpha_q|^2 + |\beta_q|^2}{2(m_{X,q}^2-m_\chi^2)}
  .
\end{align}
In the chiral limit, the $X$ vector bosons that couple to left- and 
right-handed quarks are in general distinct particles. In the absence of 
tuning, it is therefore natural for one of these two contributions to 
dominate. For the left-handed contribution, if the dark matter particle is 
a SM singlet then SM SU(2$)_{\rm L}$ symmetry implies that $d_u = d_d$.  
Further, flavor constraints require $d_d = d_s$. Then the numerical values 
of $\Delta_u$, $\Delta_d$ and $\Delta_s$ imply that there are large 
cancellations among the contributions of the different left-handed quarks 
to the WIMP-nucleon cross section, which is therefore somewhat suppressed. 
For the right-handed contribution, this cancellation can be avoided if 
$d_u \gg d_d ( = d_s)$, or vice versa.

In figures \ref{fig:lhcu} and \ref{fig:lhc} we have plotted the range
of values of the $X$ masses that would lead to a signal at IceCube,
assuming annihilation to the $W^+W^-$ final state (in general through
a distinct set of mediators). In Fig \ref{fig:lhcu} we have set
$d_d=d_s=0$ and $d_u\neq0$, with $\alpha_u = -\beta_u = \frac12$,
corresponding to one natural possibility for the contribution from
right-handed quarks.  In Fig \ref{fig:lhc}, on the other hand, we have
set $d_u = d_d = d_s$, with $\alpha_q = \beta_q = \frac12$
corresponding to the contribution from left-handed quarks.  We see
that away from the resonance region at $m_{\chi} = m_{X}$, the colored
vector boson masses lie at a TeV or below in most of the parameter
space. They are therefore kinematically accessible to the LHC, and can
be pair-produced through strong interactions. They then decay, leading
to a jets + missing energy signal. Recent model-independent studies of
dark matter signals at the LHC involving jets + missing energy have
been performed, for example, in \cite{Alwall:2008va,Izaguirre:2010nj},
\cite{Cao:2009uw}, \cite{Beltran:2010ww}, and the results look
promising.

\subsubsection* {s- and u-channel scalar exchange}
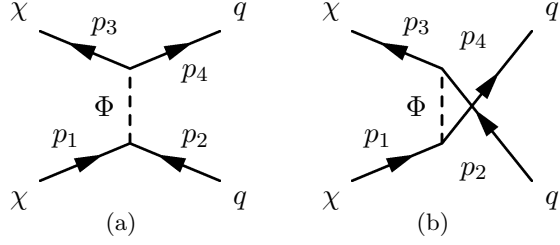
\begin{figure}[h]
  \begin{center}
    \begin{fmffile}{mc}
      \subfloat[]{
      \label{fig:mca}
      \begin{fmfgraph*}(30,20)
	\fmfleftn{ic}{2} \fmfrightn{oc}{2}
	\fmflabel{$\chi$}{ic1}
	\fmflabel{$\chi$}{ic2}
	\fmflabel{$q$}{oc1}
	\fmflabel{$q$}{oc2}
	\fmf{fermion, label=$p_1$}{ic1,vc1}
	\fmf{fermion, label=$p_3$}{vc2,ic2}
	\fmf{dashes, label=$\Phi$}{vc1,vc2}
	\fmf{fermion, label=$p_2$}{oc1,vc1}
	\fmf{fermion, label=$p_4$}{vc2,oc2}
      \end{fmfgraph*}}
      \hspace{0.3in}
      \subfloat[]{
      \label{fig:mcb}
      \begin{fmfgraph*}(30,20)
	\fmfleftn{id}{2} \fmfrightn{od}{2}
	\fmflabel{$\chi$}{id1}
	\fmflabel{$\chi$}{id2}
	\fmflabel{$q$}{od1}
	\fmflabel{$q$}{od2}
	\fmf{fermion, label=$p_1$}{id1,vd1}
	\fmf{fermion, label=$p_3$}{vd2,id2}
	\fmf{dashes, label=$\Phi$}{vd1,vd2}
	\fmf{phantom, label=$p_2$,l.s=left,l.d=0.01mm}{od1,vd1}
	\fmf{phantom, label=$p_4$,l.s=left,l.d=0.01mm}{vd2,od2}
	\fmf{fermion,tension=0}{od1,vd2}
	\fmf{fermion,tension=0}{vd1,od2}
      \end{fmfgraph*}}
    \end{fmffile}
  \end{center}
  \caption{Majorana dark matter scattering through scalar exchange}
  \label{fig:mc}
\end{figure}
Finally we consider scattering through s- and u-channel exchange of a
colored scalar $\Phi$.
The Lagrangian corresponding to this process is
\begin{align}
  \mathcal{L}
  &=
  |\partial \Phi|^2-m_\Phi^2|\Phi|^2
  -\bar{\chi}(\alpha-\beta\gamma^5)q \Phi
  -\bar{q}(\alpha^*+\beta^*\gamma^5)\chi \Phi^{\dagger}
  .
\end{align}
The coefficient of the relevant operator is given by,
\begin{align}
  d_q &= 
  \frac{|\alpha_q|^2 + |\beta_q|^2}{4(m_{\Phi,q}^2-m_{\chi}^2)}
  .
\end{align}
As explained earlier, in the chiral limit the mediators of left- and 
right-handed interactions are distinct.  In figures \ref{fig:lhcu} and 
\ref{fig:lhc} we have shown the range of values of the $\Phi$ mass that 
would lead to a signal at IceCube. In Fig \ref{fig:lhcu} we have set 
$d_u\neq0$ and $d_d=d_s=0$, with $\alpha_u = -\beta_u =\frac12$, 
corresponding to the right-handed contribution. In Fig \ref{fig:lhc} we 
have set $d_u = d_d = d_s$ with $\alpha_q = \beta_q =\frac12$, 
corresponding to the left-handed contribution.  We see that away from the 
resonance region the $\Phi$ masses lie below a TeV, which is promising for 
the LHC.

\afterpage{\clearpage}
\subsection{Vector Dark Matter}
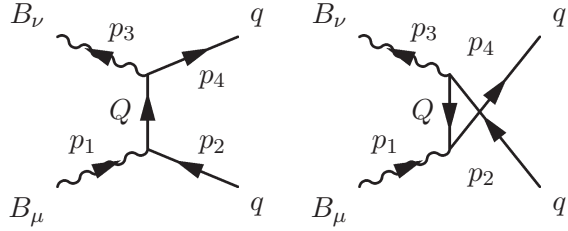
\begin{figure}[h]
  \begin{center}
    \vspace{0.3in}
    \begin{fmffile}{v}
      \begin{fmfgraph*}(30,20)
	\fmfleftn{ic}{2} \fmfrightn{oc}{2}
	\fmflabel{$B_\mu$}{ic1}
	\fmflabel{$B_\nu$}{ic2}
	\fmflabel{$q$}{oc1}
	\fmflabel{$q$}{oc2}
	\fmf{boson}{ic1,vc1}
	\fmf{phantom_arrow, label=$p_1$, tension=0}{ic1,vc1}
	\fmf{boson}{ic2,vc2}
	\fmf{phantom_arrow, label=$p_3$, tension=0}{vc2,ic2}
	\fmf{fermion,label=$Q$}{vc1,vc2}
	\fmf{fermion, label=$p_2$}{oc1,vc1}
	\fmf{fermion, label=$p_4$}{vc2,oc2}
      \end{fmfgraph*}
      \hspace{0.3in}
      \begin{fmfgraph*}(30,20)
	\fmfleftn{id}{2} \fmfrightn{od}{2}
	\fmflabel{$B_\mu$}{id1}
	\fmflabel{$B_\nu$}{id2}
	\fmflabel{$q$}{od1}
	\fmflabel{$q$}{od2}
	\fmf{boson}{id1,vd1}
	\fmf{boson}{vd2,id2}
	\fmf{phantom_arrow, label=$p_1$, tension=0}{id1,vd1}
	\fmf{phantom_arrow, label=$p_3$, tension=0}{vd2,id2}
	\fmf{fermion, label=$Q$,l.s=right}{vd2,vd1}
	\fmf{phantom, label=$p_2$,l.s=left,l.d=0.01mm}{od1,vd1}
	\fmf{phantom, label=$p_4$,l.s=left,l.d=0.01mm}{vd2,od2}
	\fmf{fermion,tension=0}{od1,vd2}
	\fmf{fermion,tension=0}{vd1,od2}
      \end{fmfgraph*}
    \end{fmffile}
  \end{center}
  \caption{Vector dark matter scattering through s and 
           u-channel fermion exchange.}
	   \label{fig:v}
    \vspace{0.3in}
\end{figure}

\begin{figure}[htp]
  \begin{center}
    \subfloat[]{
    \includegraphics[scale=1.0]{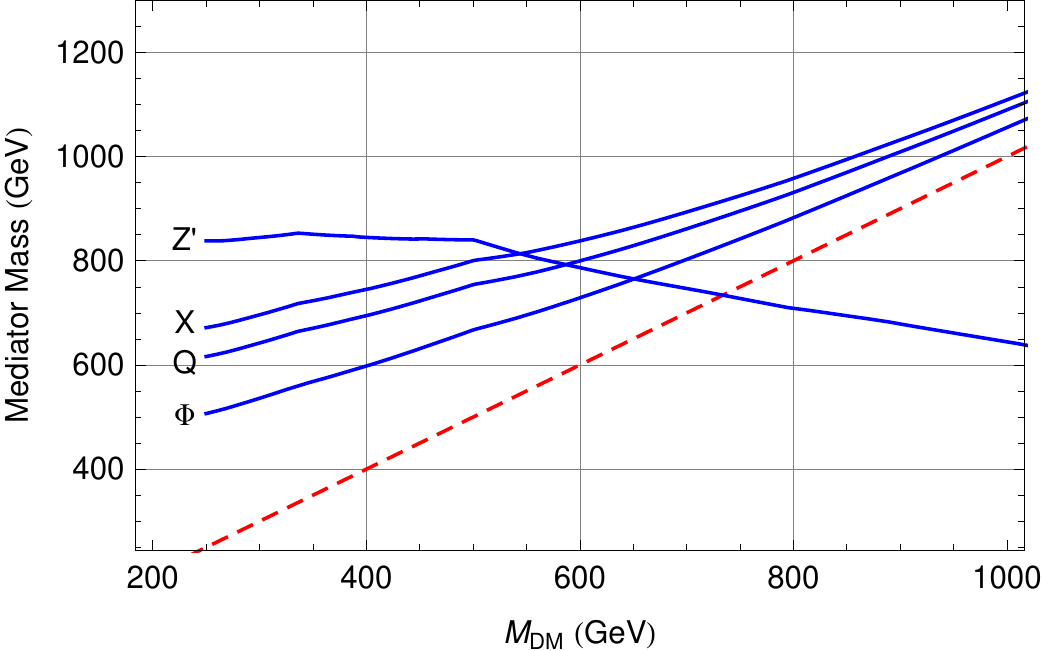}
    \label{fig:lhca}
    }
    \vspace{0.3in}
    \\
    \subfloat[]{
    \includegraphics[scale=1.0]{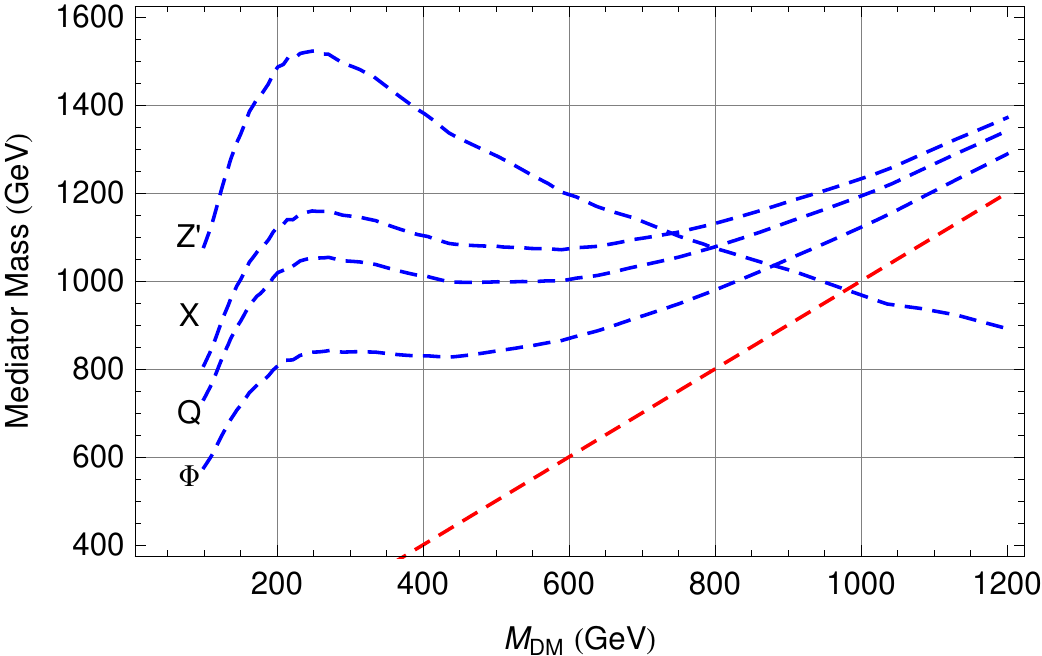}
  \label{fig:lhcb}
    }
  \end{center}
  \caption{Estimates for the mediator masses if IceCube sees a signal
  near the present bound (above) or at the 5-year sensitivity (below).
  The colored mediators $X,\Phi$
  and $Q$ are assumed to couple only to up-type quarks, while the
  charges of the $Z'$ are assumed to be proportional to the charges of
  the SM $Z$. The masses of $X,\Phi$ and $Q$ must lie above the red
  dashed line, which corresponds to where the mediator mass is equal
  to the dark matter mass. The threshold energy was chosen to be
  $E_{th}=10$ GeV.
  }
  \label{fig:lhcu}
\end{figure}
\begin{figure}[htp]
  \begin{center}
    \subfloat[]{
    \includegraphics[scale=1.0]{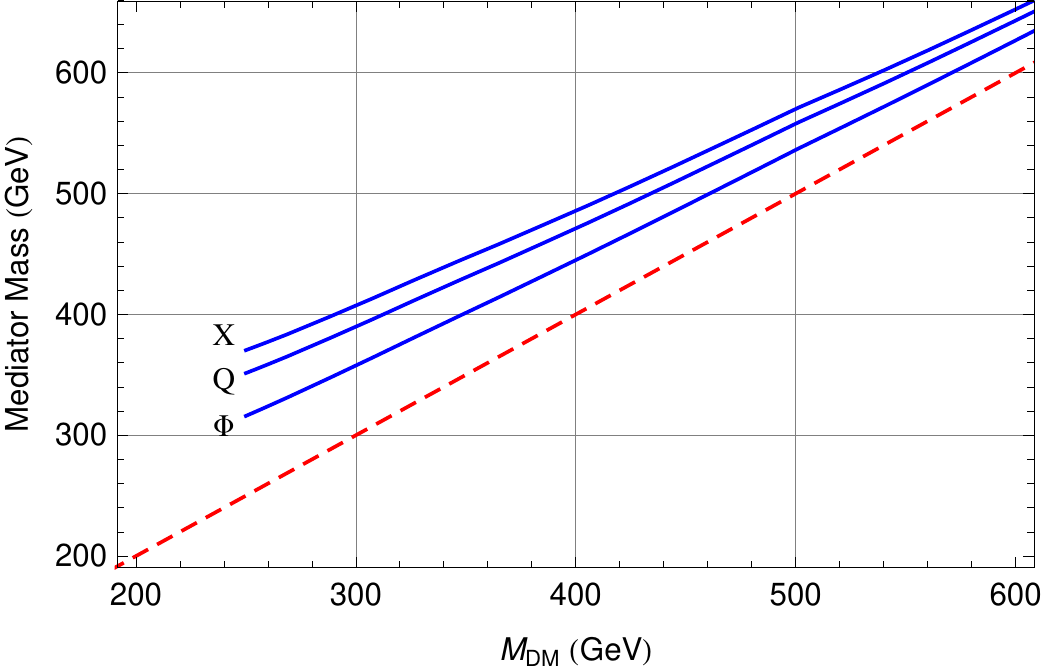}
  \label{fig:lhcc}
    }
    \vspace{0.3in}
    \\
    \subfloat[]{
    \includegraphics[scale=1.0]{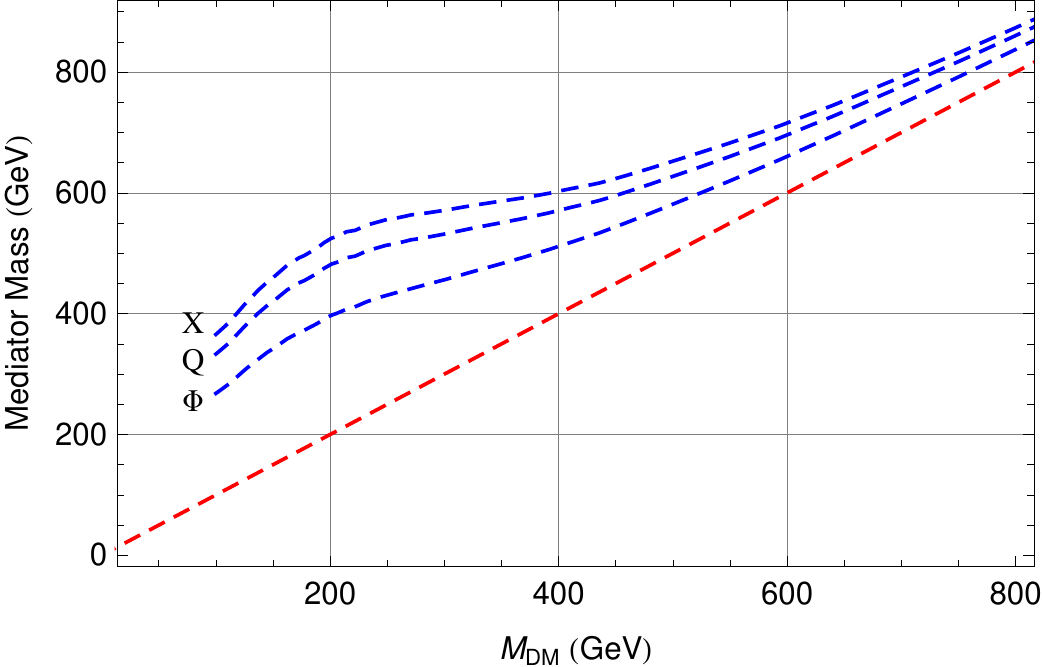}
  \label{fig:lhcd}
    }
  \end{center}
  \caption{Estimates for the colored mediator masses if IceCube sees a
  signal near the present bound (above) or at the 5-year sensitivity
  (below), assuming equal couplings to up- and down-type quarks. The
  signal in neutrino telescopes is suppressed due to
  cancellations arising from quark spin fractions, leading to lower
  values of the mediator masses.
   The threshold energy was chosen to be
  $E_{th}=10$ GeV.
  }
  \label{fig:lhc}
\end{figure}
The effective operator that generates spin-dependent scattering 
in the case that dark matter is a real vector field $B_{\mu}$ has the form
\begin{align}
  \mathcal{L}
  &=  b_q (\partial_\sigma B_\mu) B_\nu \bar{q} \gamma_\alpha \gamma^5 q
  \epsilon^{\sigma\mu\nu\alpha}
  .
\end{align}
The corresponding WIMP-nucleus cross section is then
\begin{align}
  \sigma_0
  &=
  \frac{8m_{\chi}^2 m_N^2}{3\pi \left(m_{\chi} + m_N \right)^2}
  \left[\sum_{q=u,d,s}{b_q\lambda_q}\right]^2
  J_N(J_N+1)
  .
\end{align}
This operator can be generated by the exchange of a heavy colored fermion 
$Q$ in the s- and u-channels, as shown in Fig \ref{fig:v}. The relevant 
part of the Lagrangian takes the form
\begin{align}
  \mathcal{L}
  &=
  \bar{Q}(i\slashed{\partial}-m_Q)Q
  +\bar{q}\gamma^{\mu}(\alpha-\beta\gamma^5)Q\;B_{\mu}
  +\bar{Q}\gamma^{\mu}(\alpha^*-\beta^*\gamma^5)q\;B_{\mu}
  .
\end{align}

We find that 
\begin{align}
  b_q = \frac{|\alpha_q|^2 + |\beta_q|^2}{(m_{Q,q}^2-m_B^2)}
  .
\end{align}
We seek to explore the range of masses of the mediator $Q$ and the dark 
matter particle $B_{\mu}$ that give rise to a signal at IceCube, and the 
resulting implications for the LHC. As in the cases of $X$ and $\Phi$, in 
the chiral limit the mediators $Q$ that couple to left-and right-handed 
quarks are in general different particles, and the mediators corresponding 
to different flavors are also distinct. Flavor constraints are satisfied 
provided the mediators associated with different flavors are degenerate, 
and their couplings are flavor diagonal. For concreteness we employ 
exactly the same conventions as earlier. Specifically, we first consider 
$b_u \neq 0$, $b_d = b_s=0$ with $\alpha_u = -\beta_u = \frac{1}{2}$, 
corresponding to the contribution from right-handed quarks. The results 
are plotted in Fig \ref{fig:lhcu}. We then consider $b_u = b_d = b_s$ with 
$\alpha_q = \beta_q =\frac12$, corresponding to the contribution from 
left-handed quarks. The results are plotted in Fig \ref{fig:lhc}. From the 
figures, we see that a signal at current generation neutrino telescopes 
implies that the masses of the colored fermions lie at a TeV or below, 
which is within the kinematic reach of the LHC.

\section{Conclusions}

In conclusion we have used the constraints from direct detection 
experiments to establish a model-independent upper bound on dark matter 
event rates in neutrino telescopes, which applies to any elastic dark 
matter model. The strength of the bound depends on whether the 
WIMP-nucleon cross section is spin-independent or spin-dependent. While 
the spin-dependent bound is not competitive with present limits from 
neutrino telescopes, the spin-independent bound is much tighter, and is 
comparable to current experimental bounds. As direct detection experiments 
improve, these bounds are expected to get even tighter.

If the observed event rate in neutrino telescopes exceeds the bound corresponding 
to spin-independent interactions, we can immediately infer that WIMP-nucleon 
scattering is dominated by spin-dependent interactions. The dark matter 
candidates that naturally have this property have been classified, and are known 
to be either Majorana fermions or real vector bosons, so that the dark matter 
particle is its own anti-particle. Furthermore, it is known that each such 
candidate theory predicts either new particles at the weak scale with SM quantum 
numbers, or a new $Z'$ gauge boson with mass close to the weak scale. We have 
studied the parameter space of these theories which leads to an observable signal 
at IceCube, and found that while the scenario with a new $Z'$ gauge boson is 
somewhat disfavored by current experimental constraints, the alternate scenario 
is completely viable. The masses of the new particles are favored to lie at 
energy scales that are kinematically accessible to the LHC.

\acknowledgments
We would like to thank Lian-Tao Wang for comments. PA, ZC and RKM are 
supported by the National Science Foundation under grant number 
PHY-0801323. CK is supported by the Department of Energy under grant 
DE-FG0296ER50959.

\bibliography{references}

\providecommand{\href}[2]{#2}\begingroup\raggedright\begin{thebibliography}{10}

\bibitem{Ellis:1983ew}
J.~R. Ellis, J.~S. Hagelin, D.~V. Nanopoulos, K.~A. Olive, and M.~Srednicki,
  {\it {Supersymmetric relics from the big bang}},  {\em Nucl. Phys.} {\bf
  B238} (1984) 453--476.

\bibitem{Griest:1988ma}
K.~Griest, {\it {Cross sections, relic abundance, and detection rates for
  neutralino dark matter}},  {\em Phys. Rev.} {\bf D38} (1988) 2357.

\bibitem{Kolb:1983fm}
E.~W. Kolb and R.~Slansky, {\it {Dimensional Reduction in the Early Universe:
  Where Have the Massive Particles Gone?}},  {\em Phys. Lett.} {\bf B135}
  (1984) 378.

\bibitem{Servant:2002aq}
G.~Servant and T.~M.~P. Tait, {\it {Is the lightest Kaluza-Klein particle a
  viable dark matter candidate?}},  {\em Nucl. Phys.} {\bf B650} (2003)
  391--419, [\href{http://xxx.lanl.gov/abs/hep-ph/0206071}{{\tt
  hep-ph/0206071}}].

\bibitem{BirkedalHansen:2003mpa}
A.~Birkedal-Hansen and J.~G. Wacker, {\it {Scalar dark matter from theory
  space}},  {\em Phys. Rev.} {\bf D69} (2004) 065022,
  [\href{http://xxx.lanl.gov/abs/hep-ph/0306161}{{\tt hep-ph/0306161}}].

\bibitem{Hubisz:2004ft}
J.~Hubisz and P.~Meade, {\it {Phenomenology of the littlest Higgs with
  T-parity}},  {\em Phys. Rev.} {\bf D71} (2005) 035016,
  [\href{http://xxx.lanl.gov/abs/hep-ph/0411264}{{\tt hep-ph/0411264}}].

\bibitem{Birkedal:2006fz}
A.~Birkedal, A.~Noble, M.~Perelstein, and A.~Spray, {\it {Little Higgs dark
  matter}},  {\em Phys. Rev.} {\bf D74} (2006) 035002,
  [\href{http://xxx.lanl.gov/abs/hep-ph/0603077}{{\tt hep-ph/0603077}}].

\bibitem{Dolle:2007ce}
E.~M. Dolle and S.~Su, {\it {Dark Matter in the Left Right Twin Higgs Model}},
  {\em Phys. Rev.} {\bf D77} (2008) 075013,
  [\href{http://xxx.lanl.gov/abs/0712.1234}{{\tt arXiv:0712.1234}}].

\bibitem{Silk:1985ax}
J.~Silk, K.~A. Olive, and M.~Srednicki, {\it {The photino, the sun, and
  high-energy neutrinos}},  {\em Phys. Rev. Lett.} {\bf 55} (1985) 257--259.

\bibitem{Srednicki:1986vj}
M.~Srednicki, K.~A. Olive, and J.~Silk, {\it {High-Energy Neutrinos from the
  Sun and Cold Dark Matter}},  {\em Nucl. Phys.} {\bf B279} (1987) 804.

\bibitem{Bergstrom:1996kp}
L.~Bergstrom, J.~Edsjo, and P.~Gondolo, {\it {Indirect neutralino detection
  rates in neutrino telescopes}},  {\em Phys. Rev.} {\bf D55} (1997)
  1765--1770, [\href{http://xxx.lanl.gov/abs/hep-ph/9607237}{{\tt
  hep-ph/9607237}}].

\bibitem{Bergstrom:1998xh}
L.~Bergstrom, J.~Edsjo, and P.~Gondolo, {\it {Indirect detection of dark matter
  in km-size neutrino telescopes}},  {\em Phys. Rev.} {\bf D58} (1998) 103519,
  [\href{http://xxx.lanl.gov/abs/hep-ph/9806293}{{\tt hep-ph/9806293}}].

\bibitem{Barger:2001ur}
V.~D. Barger, F.~Halzen, D.~Hooper, and C.~Kao, {\it {Indirect search for
  neutralino dark matter with high energy neutrinos}},  {\em Phys. Rev.} {\bf
  D65} (2002) 075022, [\href{http://xxx.lanl.gov/abs/hep-ph/0105182}{{\tt
  hep-ph/0105182}}].

\bibitem{Barger:2007xf}
V.~Barger, W.-Y. Keung, G.~Shaughnessy, and A.~Tregre, {\it {High energy
  neutrinos from neutralino annihilations in the Sun}},  {\em Phys. Rev.} {\bf
  D76} (2007) 095008, [\href{http://xxx.lanl.gov/abs/0708.1325}{{\tt
  arXiv:0708.1325}}].

\bibitem{Bertin:2002ky}
V.~Bertin, E.~Nezri, and J.~Orloff, {\it {Neutrino indirect detection of
  neutralino dark matter in the CMSSM}},  {\em Eur. Phys. J.} {\bf C26} (2002)
  111--124, [\href{http://xxx.lanl.gov/abs/hep-ph/0204135}{{\tt
  hep-ph/0204135}}].

\bibitem{Bertin:2002sq}
V.~Bertin, E.~Nezri, and J.~Orloff, {\it {Neutralino dark matter beyond CMSSM
  universality}},  {\em JHEP} {\bf 02} (2003) 046,
  [\href{http://xxx.lanl.gov/abs/hep-ph/0210034}{{\tt hep-ph/0210034}}].

\bibitem{Hooper:2003ka}
D.~Hooper and L.-T. Wang, {\it {Direct and indirect detection of neutralino
  dark matter in selected supersymmetry breaking scenarios}},  {\em Phys. Rev.}
  {\bf D69} (2004) 035001, [\href{http://xxx.lanl.gov/abs/hep-ph/0309036}{{\tt
  hep-ph/0309036}}].

\bibitem{Jungman:1995df}
G.~Jungman, M.~Kamionkowski, and K.~Griest, {\it {Supersymmetric dark matter}},
   {\em Phys. Rept.} {\bf 267} (1996) 195--373,
  [\href{http://xxx.lanl.gov/abs/hep-ph/9506380}{{\tt hep-ph/9506380}}].

\bibitem{Cheng:2002ej}
H.-C. Cheng, J.~L. Feng, and K.~T. Matchev, {\it {Kaluza-Klein dark matter}},
  {\em Phys. Rev. Lett.} {\bf 89} (2002) 211301,
  [\href{http://xxx.lanl.gov/abs/hep-ph/0207125}{{\tt hep-ph/0207125}}].

\bibitem{Hooper:2002gs}
D.~Hooper and G.~D. Kribs, {\it {Probing Kaluza-Klein dark matter with neutrino
  telescopes}},  {\em Phys. Rev.} {\bf D67} (2003) 055003,
  [\href{http://xxx.lanl.gov/abs/hep-ph/0208261}{{\tt hep-ph/0208261}}].

\bibitem{Dobrescu:2007ec}
B.~A. Dobrescu, D.~Hooper, K.~Kong, and R.~Mahbubani, {\it {Spinless photon
  dark matter from two universal extra dimensions}},  {\em JCAP} {\bf 0710}
  (2007) 012, [\href{http://xxx.lanl.gov/abs/0706.3409}{{\tt
  arXiv:0706.3409}}].

\bibitem{Blennow:2009ag}
M.~Blennow, H.~Melbeus, and T.~Ohlsson, {\it {Neutrinos from Kaluza-Klein dark
  matter in the Sun}},  \href{http://xxx.lanl.gov/abs/0910.1588}{{\tt
  arXiv:0910.1588}}.

\bibitem{Flacke:2009eu}
T.~Flacke, A.~Menon, D.~Hooper, and K.~Freese, {\it {Kaluza-Klein Dark Matter
  And Neutrinos From Annihilation In The Sun}},
  \href{http://xxx.lanl.gov/abs/0908.0899}{{\tt arXiv:0908.0899}}.

\bibitem{Hooper:2005fj}
D.~Hooper and G.~Servant, {\it {Indirect detection of Dirac right-handed
  neutrino dark matter}},  {\em Astropart. Phys.} {\bf 24} (2005) 231--246,
  [\href{http://xxx.lanl.gov/abs/hep-ph/0502247}{{\tt hep-ph/0502247}}].

\bibitem{Perelstein:2006bq}
M.~Perelstein and A.~Spray, {\it {Indirect detection of little Higgs dark
  matter}},  {\em Phys. Rev.} {\bf D75} (2007) 083519,
  [\href{http://xxx.lanl.gov/abs/hep-ph/0610357}{{\tt hep-ph/0610357}}].

\bibitem{Agrawal:2008xz}
P.~Agrawal, E.~M. Dolle, and C.~A. Krenke, {\it {Signals of Inert Doublet Dark
  Matter in Neutrino Telescopes}},  {\em Phys. Rev.} {\bf D79} (2009) 015015,
  [\href{http://xxx.lanl.gov/abs/0811.1798}{{\tt arXiv:0811.1798}}].

\bibitem{Andreas:2009hj}
S.~Andreas, M.~H.~G. Tytgat, and Q.~Swillens, {\it {Neutrinos from Inert
  Doublet Dark Matter}},  {\em JCAP} {\bf 0904} (2009) 004,
  [\href{http://xxx.lanl.gov/abs/0901.1750}{{\tt arXiv:0901.1750}}].

\bibitem{Andreas:2009jp}
S.~Andreas, {\it {Neutrino signature of Inert Doublet Dark Matter}},
  \href{http://xxx.lanl.gov/abs/0911.0540}{{\tt arXiv:0911.0540}}.

\bibitem{Allahverdi:2009se}
R.~Allahverdi, S.~Bornhauser, B.~Dutta, and K.~Richardson-McDaniel, {\it
  {Prospects for Indirect Detection of Sneutrino Dark Matter with IceCube}},
  {\em Phys. Rev.} {\bf D80} (2009) 055026,
  [\href{http://xxx.lanl.gov/abs/0907.1486}{{\tt arXiv:0907.1486}}].

\bibitem{Kamionkowski:1994dp}
M.~Kamionkowski, K.~Griest, G.~Jungman, and B.~Sadoulet, {\it {Model
  independent comparison of direct versus indirect detection of supersymmetric
  dark matter}},  {\em Phys. Rev. Lett.} {\bf 74} (1995) 5174--5177,
  [\href{http://xxx.lanl.gov/abs/hep-ph/9412213}{{\tt hep-ph/9412213}}].

\bibitem{Halzen:2005ar}
F.~Halzen and D.~Hooper, {\it {Prospects for detecting dark matter with
  neutrino telescopes in light of recent results from direct detection
  experiments}},  {\em Phys. Rev.} {\bf D73} (2006) 123507,
  [\href{http://xxx.lanl.gov/abs/hep-ph/0510048}{{\tt hep-ph/0510048}}].

\bibitem{Wikstrom:2009kw}
G.~Wikstrom and J.~Edsjo, {\it {Limits on the WIMP-nucleon scattering
  cross-section from neutrino telescopes}},  {\em JCAP} {\bf 0904} (2009) 009,
  [\href{http://xxx.lanl.gov/abs/0903.2986}{{\tt arXiv:0903.2986}}].

\bibitem{Bandyopadhyay:2010zk}
A.~Bandyopadhyay, S.~Chakraborty, and D.~Majumdar, {\it {Interpreting the
  bounds on Dark Matter induced muons at Super-Kamiokande in the light of CDMS
  data}},  \href{http://xxx.lanl.gov/abs/1002.0753}{{\tt arXiv:1002.0753}}.

\bibitem{Abbasi:2009uz}
{\bf IceCube} Collaboration, R.~Abbasi {\em et.~al.}, {\it {Limits on a muon
  flux from neutralino annihilations in the Sun with the IceCube 22-string
  detector}},  {\em Phys. Rev. Lett.} {\bf 102} (2009) 201302,
  [\href{http://xxx.lanl.gov/abs/0902.2460}{{\tt arXiv:0902.2460}}].

\bibitem{Abbasi:2009vg}
{\bf IceCube} Collaboration, .~R. Abbasi, {\it {Limits on a muon flux from
  Kaluza-Klein dark matter annihilations in the Sun from the IceCube 22-string
  detector}},  \href{http://xxx.lanl.gov/abs/0910.4480}{{\tt arXiv:0910.4480}}.

\bibitem{Agrawal:2010fh}
P.~Agrawal, Z.~Chacko, C.~Kilic, and R.~K. Mishra, {\it {Direct Detection
  Constraints on Dark Matter Event Rates in Neutrino Telescopes, and Collider
  Implications}},  \href{http://xxx.lanl.gov/abs/1003.1912}{{\tt
  arXiv:1003.1912}}.

\bibitem{Ahmed:2009zw}
{\bf The CDMS-II} Collaboration, Z.~Ahmed {\em et.~al.}, {\it {Results from the
  Final Exposure of the CDMS II Experiment}},
  \href{http://xxx.lanl.gov/abs/0912.3592}{{\tt arXiv:0912.3592}}.

\bibitem{Angle:2008we}
J.~Angle {\em et.~al.}, {\it {Limits on spin-dependent WIMP-nucleon
  cross-sections from the XENON10 experiment}},  {\em Phys. Rev. Lett.} {\bf
  101} (2008) 091301, [\href{http://xxx.lanl.gov/abs/0805.2939}{{\tt
  arXiv:0805.2939}}].

\bibitem{Archambault:2009sm}
S.~Archambault {\em et.~al.}, {\it {Dark Matter Spin-Dependent Limits for WIMP
  Interactions on 19-F by PICASSO}},  {\em Phys. Lett.} {\bf B682} (2009)
  185--192, [\href{http://xxx.lanl.gov/abs/0907.0307}{{\tt arXiv:0907.0307}}].

\bibitem{Kim:2008zzn}
{\bf KIMS} Collaboration, S.~K. Kim, {\it {New results from the KIMS
  experiment}},  {\em J. Phys. Conf. Ser.} {\bf 120} (2008) 042021.

\bibitem{Gould1992}
A.~Gould, {\it {Cosmological density of WIMPs from solar and terrestrial
  annihilations}},  {\em Astrophys. J.} {\bf 388} (1992) 338--344.

\bibitem{Gondolo:2004sc}
P.~Gondolo {\em et.~al.}, {\it {DarkSUSY: Computing supersymmetric dark matter
  properties numerically}},  {\em JCAP} {\bf 0407} (2004) 008,
  [\href{http://xxx.lanl.gov/abs/astro-ph/0406204}{{\tt astro-ph/0406204}}].

\bibitem{Dutta:2000hh}
S.~I. Dutta, M.~H. Reno, I.~Sarcevic, and D.~Seckel, {\it {Propagation of muons
  and taus at high energies}},  {\em Phys. Rev.} {\bf D63} (2001) 094020,
  [\href{http://xxx.lanl.gov/abs/hep-ph/0012350}{{\tt hep-ph/0012350}}].

\bibitem{Gandhi:1995tf}
R.~Gandhi, C.~Quigg, M.~H. Reno, and I.~Sarcevic, {\it {Ultrahigh-energy
  neutrino interactions}},  {\em Astropart. Phys.} {\bf 5} (1996) 81--110,
  [\href{http://xxx.lanl.gov/abs/hep-ph/9512364}{{\tt hep-ph/9512364}}].

\bibitem{Kopp:2009et}
J.~Kopp, V.~Niro, T.~Schwetz, and J.~Zupan, {\it {DAMA/LIBRA and leptonically
  interacting Dark Matter}},  {\em Phys. Rev.} {\bf D80} (2009) 083502,
  [\href{http://xxx.lanl.gov/abs/0907.3159}{{\tt arXiv:0907.3159}}].

\bibitem{Aprile:2009zzb}
E.~Aprile, {\it {The XENON100 dark matter experiment}},  {\em AIP Conf. Proc.}
  {\bf 1115} (2009) 355--360.

\bibitem{Fiorucci:2009ak}
S.~Fiorucci {\em et.~al.}, {\it {Status of the LUX Dark Matter Search}},
  \href{http://xxx.lanl.gov/abs/0912.0482}{{\tt arXiv:0912.0482}}.

\bibitem{Desai:2007ra}
{\bf Super-Kamiokande} Collaboration, S.~Desai {\em et.~al.}, {\it {Study of
  TeV Neutrinos with Upward Showering Muons in Super-Kamiokande}},  {\em
  Astropart. Phys.} {\bf 29} (2008) 42--54,
  [\href{http://xxx.lanl.gov/abs/0711.0053}{{\tt arXiv:0711.0053}}].

\bibitem{clercq2008}
{\bf IceCube} Collaboration, C.~de~Clercq, {\it {Search for Dark Matter with
  the AMANDA and IceCube neutrino detectors (20'+5')}},  in {\em Identification
  of dark matter}, 2008.

\bibitem{sommerfeld1931}
A.~Sommerfeld, {\it {ber die Beugung und Bremsung der Elektronen}},  {\em
  {Annalen der Physik}} {\bf 403} (1931), no.~3 257--330.

\bibitem{Cirelli:2007xd}
M.~Cirelli, A.~Strumia, and M.~Tamburini, {\it {Cosmology and Astrophysics of
  Minimal Dark Matter}},  {\em Nucl. Phys.} {\bf B787} (2007) 152--175,
  [\href{http://xxx.lanl.gov/abs/0706.4071}{{\tt arXiv:0706.4071}}].

\bibitem{ArkaniHamed:2008qn}
N.~Arkani-Hamed, D.~P. Finkbeiner, T.~R. Slatyer, and N.~Weiner, {\it {A Theory
  of Dark Matter}},  {\em Phys. Rev.} {\bf D79} (2009) 015014,
  [\href{http://xxx.lanl.gov/abs/0810.0713}{{\tt arXiv:0810.0713}}].

\bibitem{Bedaque:2009ri}
P.~F. Bedaque, M.~I. Buchoff, and R.~K. Mishra, {\it {Sommerfeld enhancement
  from Goldstone pseudo-scalar exchange}},  {\em JHEP} {\bf 11} (2009) 046,
  [\href{http://xxx.lanl.gov/abs/0907.0235}{{\tt arXiv:0907.0235}}].

\bibitem{TuckerSmith:2001hy}
D.~Tucker-Smith and N.~Weiner, {\it {Inelastic dark matter}},  {\em Phys. Rev.}
  {\bf D64} (2001) 043502, [\href{http://xxx.lanl.gov/abs/hep-ph/0101138}{{\tt
  hep-ph/0101138}}].

\bibitem{TuckerSmith:2004jv}
D.~Tucker-Smith and N.~Weiner, {\it {The status of inelastic dark matter}},
  {\em Phys. Rev.} {\bf D72} (2005) 063509,
  [\href{http://xxx.lanl.gov/abs/hep-ph/0402065}{{\tt hep-ph/0402065}}].

\bibitem{Cui:2009xq}
Y.~Cui, D.~E. Morrissey, D.~Poland, and L.~Randall, {\it {Candidates for
  Inelastic Dark Matter}},  {\em JHEP} {\bf 05} (2009) 076,
  [\href{http://xxx.lanl.gov/abs/0901.0557}{{\tt arXiv:0901.0557}}].

\bibitem{Feldstein:2009tr}
B.~Feldstein, A.~L. Fitzpatrick, and E.~Katz, {\it {Form Factor Dark Matter}},
  \href{http://xxx.lanl.gov/abs/0908.2991}{{\tt arXiv:0908.2991}}.

\bibitem{Chang:2009yt}
S.~Chang, A.~Pierce, and N.~Weiner, {\it {Momentum Dependent Dark Matter
  Scattering}},  \href{http://xxx.lanl.gov/abs/0908.3192}{{\tt
  arXiv:0908.3192}}.

\bibitem{Bai:2009cd}
Y.~Bai and P.~J. Fox, {\it {Resonant Dark Matter}},  {\em JHEP} {\bf 11} (2009)
  052, [\href{http://xxx.lanl.gov/abs/0909.2900}{{\tt arXiv:0909.2900}}].

\bibitem{Nussinov:2009ft}
S.~Nussinov, L.-T. Wang, and I.~Yavin, {\it {Capture of Inelastic Dark Matter
  in the Sun}},  {\em JCAP} {\bf 0908} (2009) 037,
  [\href{http://xxx.lanl.gov/abs/0905.1333}{{\tt arXiv:0905.1333}}].

\bibitem{Menon:2009qj}
A.~Menon, R.~Morris, A.~Pierce, and N.~Weiner, {\it {Capture and Indirect
  Detection of Inelastic Dark Matter}},
  \href{http://xxx.lanl.gov/abs/0905.1847}{{\tt arXiv:0905.1847}}.

\bibitem{Ellis:2000ds}
J.~R. Ellis, A.~Ferstl, and K.~A. Olive, {\it {Re-evaluation of the elastic
  scattering of supersymmetric dark matter}},  {\em Phys. Lett.} {\bf B481}
  (2000) 304--314, [\href{http://xxx.lanl.gov/abs/hep-ph/0001005}{{\tt
  hep-ph/0001005}}].

\bibitem{Mallot:1999qb}
G.~K. Mallot, {\it {The spin structure of the nucleon}},
  \href{http://xxx.lanl.gov/abs/hep-ex/9912040}{{\tt hep-ex/9912040}}.

\bibitem{Cohen:2010gj}
T.~Cohen, D.~J. Phalen, and A.~Pierce, {\it {On the Correlation Between the
  Spin-Independent and Spin- Dependent Direct Detection of Dark Matter}},
  \href{http://xxx.lanl.gov/abs/1001.3408}{{\tt arXiv:1001.3408}}.

\bibitem{Appelquist:2002mw}
T.~Appelquist, B.~A. Dobrescu, and A.~R. Hopper, {\it {Nonexotic neutral gauge
  bosons}},  {\em Phys. Rev.} {\bf D68} (2003) 035012,
  [\href{http://xxx.lanl.gov/abs/hep-ph/0212073}{{\tt hep-ph/0212073}}].

\bibitem{Amsler:2008zzb}
{\bf Particle Data Group} Collaboration, C.~Amsler {\em et.~al.}, {\it {Review
  of particle physics}},  {\em Phys. Lett.} {\bf B667} (2008) 1.

\bibitem{Alwall:2008va}
J.~Alwall, M.-P. Le, M.~Lisanti, and J.~G. Wacker, {\it {Model-Independent Jets
  plus Missing Energy Searches}},  {\em Phys. Rev.} {\bf D79} (2009) 015005,
  [\href{http://xxx.lanl.gov/abs/0809.3264}{{\tt arXiv:0809.3264}}].

\bibitem{Izaguirre:2010nj}
E.~Izaguirre, M.~Manhart, and J.~G. Wacker, {\it {Bigger, Better, Faster, More
  at the LHC}},  \href{http://xxx.lanl.gov/abs/1003.3886}{{\tt
  arXiv:1003.3886}}.

\bibitem{Cao:2009uw}
Q.-H. Cao, C.-R. Chen, C.~S. Li, and H.~Zhang, {\it {Effective Dark Matter
  Model: Relic density, CDMS II, Fermi LAT and LHC}},
  \href{http://xxx.lanl.gov/abs/0912.4511}{{\tt arXiv:0912.4511}}.

\bibitem{Beltran:2010ww}
M.~Beltran, D.~Hooper, E.~W. Kolb, Z.~A.~C. Krusberg, and T.~M.~P. Tait, {\it
  {Maverick dark matter at colliders}},
  \href{http://xxx.lanl.gov/abs/1002.4137}{{\tt arXiv:1002.4137}}.

\end{thebibliography}\endgroup
\bibliographystyle{JHEP}    % (uses file "plain.bst")

\end{document}